\documentclass[10pt,journal,onecolumn]{IEEEtran}
\usepackage{ifpdf}
\usepackage[cmex10]{amsmath}
\usepackage{algorithmic}
\usepackage{array}
\usepackage{fixltx2e}
\usepackage{stfloats}
\usepackage{url}
\usepackage{color}
\usepackage{algorithm}
\usepackage{algorithmic}
\usepackage{textcmds}
\usepackage{amsthm}
\theoremstyle{definition}

\newtheorem{theorem}{Theorem}[section]

\newtheorem{proposition}[theorem]{Proposition}

\ifCLASSOPTIONcompsoc
  \usepackage[nocompress]{cite}
\else
  \usepackage{cite}
\fi

\ifCLASSINFOpdf
   \usepackage[pdftex]{graphicx}
\else
  \usepackage[dvips]{graphicx}
\fi

\ifCLASSOPTIONcompsoc
  \usepackage[caption=false,font=footnotesize,labelfont=sf,textfont=sf]{subfig}
\else
  \usepackage[caption=false,font=footnotesize]{subfig}
\fi

\begin{document}

\title{Measuring spatial uniformity with the hypersphere chord length distribution}

\author{Panagiotis~Sidiropoulos,~\IEEEmembership{Member,~IEEE}%
\IEEEcompsocitemizethanks{\IEEEcompsocthanksitem P.~Sidiropoulos is with the Mullard Space Science Laboratory / University College London (UCL), UK.\protect ~email: p.sidiropoulos@ucl.ac.uk.}%
}

\markboth{}%
{Measuring spatial uniformity with the hypersphere chord length distribution}
%

\IEEEtitleabstractindextext{%
\begin{abstract}

Data uniformity is a concept associated with several semantic data characteristics such as lack of features, correlation and sample bias. This article introduces a novel measure to assess data uniformity and detect uniform pointsets on high-dimensional Euclidean spaces. Spatial uniformity measure builds upon the isomorphism between hyperspherical chords and $L2$-normalised data Euclidean distances, which is implied by the fact that, in Euclidean spaces, $L2$-normalised data can be geometrically defined as points on a hypersphere. The imposed connection between the distance distribution of uniformly selected points and the hyperspherical chord length distribution is employed to quantify uniformity. More specifically,, the closed-form expression of hypersphere chord length distribution is revisited extended, before examining a few qualitative and quantitative characteristics of this distribution that can be rather straightforwardly linked to data uniformity. The experimental section includes validation in four distinct setups, thus substantiating the potential of the new uniformity measure on practical data-science applications.
\end{abstract}

}

\maketitle

\IEEEdisplaynontitleabstractindextext

\IEEEpeerreviewmaketitle

\section{Introduction}
\label{sec:introduction}

Uniformity is universally recognised across scientific domains, being used in a wide range of applications, since it is connected with several semantic data characteristics. Examples include but are not limited to (1) aggregating points in multidimensional feature space (in which case uniformity suggests the lack of distinctive features), (2) concatenating uncorrelated Gaussian variables on a single vector (which may generate uniform points on a hypersphere through normalisation \cite{ABlum16} and (3) tuning multiple hyperparameters during algorithm evaluation, in cases that the hyperparameters number prohibits the brute-force testing of all parameter combinations (so as to avoid over-representation or under-representation of regions on the hyperparameter space). On the other hand, uniformity is considered ''common knowledge'' and is rarely cited. The underlying intuitive definition of uniformity is as follows: \emph{a pointset defined on a space $S$ is uniform if-f it is the output of a stochastic process defined in $S$ in which all $p_i \in S$ have equal probability $P_i = c$ to be generated}. 

The main issue with this definition is that it imposes a large (often infinite) number of probability equalities, which are both theoretically and practically challenging to fully confirm without a priori knowledge of the stochastic process. As a result, most of the times uniformity is confirmed through \textit{reductio ad absurdum} reasoning; a set of reasonable non-uniform distributions are examined and disproven, thus implying uniformity as the only valid option. Perhaps the most typical approach is to aggregate all probability equalities to a small number of subset probability equations, based on the fact that only in a (spatial) uniform distribution the probability ratio equals to the size ratio, i.e. $P_1/P_2 = |S_1|/|S_2|$, where $P_i, ~i=\{1,2\}$ is the probability of a point generated to a subset $S_i, ~i=\{1,2\}$ of $S$, with corresponding size $|S_i|$. Especially if $S$ is segmented to a family of non-overlapping equal-sized sets $S_i, \cup S_i = S$ the absolute frequency of all $S_i$ is expected to be equal iff the point distribution is uniform. In practice, this approach suffers from three main shortcomings: (a) the optimal number of subsets as well as their boundaries is not trivial to estimate, especially in cases that unmodelled symmetry properties may cause erroneous uniform identification (b) segmenting sets becomes increasingly problematic in high-dimensional spaces due to the ''curse of dimensionality'' \cite{JLBentley75} (c) the output of the assessment is a logical variable ("true" or "false") while no quantitative evaluation is conducted. 

A more direct approach to examine uniformity is through the use of spherical harmonics \cite{TMMacRobert48}. Spherical harmonics are a complete set of orthogonal functions on the hypersphere that model both uniformity and symmetry. A spatial distribution on a hypersphere being dominated by the spherical harmonic of degree $0$ (which corresponds to the uniform part of the distribution) may be declared uniform. However, despite their elegant and mathematically rigid modelling of uniformity, the generalisation of spherical harmonics to higher dimensions greatly expands the number of spherical harmonics even of low degree. As a matter of fact, the number of spherical harmonics of degree $m$ in $N$ dimensions is $\frac{2m+N-2}{m}{N+m-3 \choose m-1}$ \cite{CRFrye14}. Hence, the number of spherical harmonics of $m$ degree is linear in $3$-dimensional space, quadratic in $4$-dimensional spaces, cubic in $5$-dimensional spaces, etc. This makes impractical the use of spherical harmonics even for small $N$ values.

The foundation of the present work is a novel uniformity definition, one that is equivalent to the ''classical'' one, but can lead to additional tools to examine uniformity: \emph{a pointset defined on a space $S$ is uniform if-f it is the output of a stochastic process in which the limit set of generated points includes an equal number of all $p_i \in S$}. In the above statement, the phrase ''limit set of generated points'' refers to a set that contains infinitely more points than $S$. The novelty of this definition is that it is based on the absolute frequency of the generated points and not the probability as the classical uniformity distribution. The two are obviously equivalent because the limit at infinity of the absolute frequency is the probability. 

The main gain is that the new definition implies a connection of the uniformity distribution with the chords connecting points of $S$. More specifically, since the limit uniform set includes an equal number of all points, the limit distribution of point distances $||p_i-p_j||, p_i,p_j \in S$ is the distribution of the chord lengths of $S$. If $||.||$ is the metric of $S$ then its chord length distribution can be examined and formalised, thus modelling the distribution that uniform point distances follow. Subsequently, the similarity of pointset distance distributions with the theoretic chord length distribution can be used to qualitatively and quantitatively assess uniformity.

This work presents such an analysis, for the special case that $S$ is a hypersphere of dimension $N$ and $||.||$ is the Euclidean distance. This case is very useful from a practical point of view because it corresponds to points normalised to have a fixed Euclidean norm (usually equal to $1$), a data structure that finds extended applications on data science. Apart from the novel uniformity definition, the main novelties of this work are:
\begin{itemize}
\item The closed-form expression and the basic properties of the hypersphere chord length distribution and a corresponding analysis for the hyper-hemisphere chord length distribution
\item The introduction of the basic principles of measuring uniformity using the hypersphere chord length distribution, including a preliminary experimental evaluation on both real and synthetic data
\item The introduction of the basic principles of detecting uniform hyperspherical subsets in high-dimensional data, including a preliminary experimental evaluation
\end{itemize}

The rest of this work is structured as follows. The related work on estimating closed-form expressions of chord length distributions is summarised on Section \ref{sec:chord_length}, while the hypersphere and hyper-hemisphere chord length distributions are presented and thoroughly examined in Section \ref{sec:Chord_length_distributions}. The theoretic analysis of how this can be used to assess uniformity and detect uniform subsets is conducted on Sections \ref{sec:point_uniform} and \ref{sec:higher_dimensions}, respectively, while the related experimental evaluation follows on Section \ref{sec:applications}. Section \ref{sec:conclusions} concludes this article.

\section{Chord Length Distributions}
\label{sec:chord_length}

The study of chord length distributions is part of stochastic geometry, a domain historically being a sparse set of intuitive mathematical puzzles (such as the Buffon's clean tile and needle problems \cite{GBuffon77}, \cite{AMathai99}), which has recently significantly advanced both theoretically and practically \cite{SNChiu07}, the latter including applications in image analysis (e.g \cite{CLacoste05}, computer vision (e.g. \cite{AJBaddeley93}), etc. Within stochastic geometry, the chord length is defined as a random variable, more specifically, the random variable that is equal to the distance $||p_i-p_j||$ of two points $p_i$, $p_j$ randomly (i.e. uniformly) selected from a space $S$. The chord length distribution models this random variable, and as already mentioned, it is also the limit distribution of the inner distances of a uniformly selected set on $S$.

Despite this association, there is not a lot of work that has been done in the direction of estimating closed-form expressions of chord length distributions. Currently, this challenging problem has found solutions in very specific cases, usually related to $2$-dimensional or $3$-dimensional spaces used in radiation research \cite{FGruy15}. Examples of shapes for which the chord length distribution is known is a regular polygon \cite{UBasel14}, a parallelogram \cite{SRen12}, a cube \cite{JPhilip07}, a hemisphere \cite{JBLangworthy89}, etc.

The literature of closed-form expressions of chord length distributions in high-dimensional spaces is even more sparse, including the chord length distribution of points inside a hypersphere \cite{JMHammersley50} (which is different than the distribution on a hypersphere that is presented here), in two adjacent unit squares \cite{VSAlagar76} the chord length distribution of N-dimensional points of variables following Gaussian distribution \cite{BThirey15} and an analysis regarding specifically the average chord length in a compact convex subset of a n-dimensional Euclidean space \cite{BBurgstaller09}. 

Characteristic of the limited interest in high-dimensional chord length distributions is the fact that while J. M. Hammersley introduced the chord length distribution of points selected within a hypersphere in 1950, the corresponding chord length distribution for points selected on a hypersphere became available on a preliminary self-printed version of this work more than $6$ decades later \cite{PSidiropoulos14ar}, based on the recently estimated closed-form expression of the surface of a hyperspherical cap  as a fraction of the total hypersphere surface \cite{SLi11}. In the present version the hypersphere chord length distribution estimation is repeated in a more compact presentation, augmented by the corresponding analysis for hyper-hemispheres. Moreover, the introduced distributions are not merely presented as mathematical achievements but are subsequently employed in a novel approach that both quantitatively and qualitatively assess spatial uniformity.

\section{Chord length distributions on the hypersphere}
\label{sec:Chord_length_distributions}

\subsection{Hypersphere chord length distribution}
\label{sec:full_distribution}

Let $p_i = \{p_{i1}~p_{i2}~p_{i3}~...~p_{iN}\}, \i \in \{1,2,...M\}$ be $M$ points selected uniformly and independently from the surface of a $N$-dimensional hypersphere 
of radius $R$, i.e., $\forall ~ i \in \{1,2,...M\}, ~p_{i1}^2 + p_{i2}^2 +...p_{iN}^2=R^2$. The pairwise Euclidean distances $d(i,j), i,j \in \{1,2,...M\}, i \neq j$ 
of $p_i$, $p_j$ generate a set $d_k$ of distances ($k~=~M(M-1)/2$). The hypersphere (or N-sphere) chord length distribution $f_N(d)$ is the distribution of $d_k$ as $k$ (i.e. $M$) tends to infinity.

If $N=2$, then the N-sphere is a circle. The circle chord length distribution is a special case, for which both the pdf ($f_2(d)$) and the cdf ($F_2(d)$) can be found in the literature (e.g. \cite{EWeisstein2}):

\begin{equation}
\label{eq:circle_pdf} f_2(d) = \frac{1}{\pi}\frac{1}{\sqrt{1-\frac{d^2}{2R^2}}}
\end{equation}

\begin{equation}
\label{eq:circle_cdf} F_2(d) = \frac{cos^{-1}(1-\frac{d^2}{2R^2})}{\pi}
\end{equation}

The estimation of the closed-form expressions for the pdf and the cdf in the general case (i.e. $f_N(d)$ and $F_N(d)$, $N \geq 2$, respectively) is assisted by the hypersphere homogeneity, i.e. the fact that the hypersphere (and its chord length distribution) is invariant to axis rotation. Therefore, the hypersphere chord length distribution can be estimated assuming that one chord end is fixed to $\{0,0,0...0,R\}$, while the other end determines the chord length. An additional consequence of the rotation invariance is that $f_N(d)$ ($F_N(d)$) is not only the asymptotic pdf (cdf) of 
$d_k$ but also the asymptotic pdf (cdf) of the distances $d(i,j), j \neq i$ from any fixed point in the point set $p_i$, i.e. that when $M$ tends to infinity each row (and column) of the distance matrix $d(i,j)$ would follow $f_N(d)$ distribution.

Assuming that one of the end points of the chord are in $p=\{0,0,0...0,R\}$, the chords of length $d$ lie on a $(N-1)$-sphere of radius $a=\sqrt{d^2-\frac{d^4}{4R^2}}$. 
This is derived by eliminating $p_{iN}$ from the N-sphere equation and the distance-from-$p$ equation ($p_{i1}^2 + p_{i2}^2 +...+(p_{iN}-R)^2=d^2$). The $(N-1)$-sphere is the intersection of the $N$-sphere with the hyperplane $L: p_{N}=R-\frac{d^2}{2R}$. Since $\frac{\partial p_{N}}{\partial d} \leq 0$, for all 
points $p'$ of the N-sphere with distance $D$ from $p$, $D < d$, $p'_{iN} > R-\frac{d^2}{2R}$ and for all points $p''$ of the N-sphere with distance $D$ from $p$, $D > d$, 
$p''_{iN} < R-\frac{d^2}{2R}$. Therefore, $L$ cuts the hypersphere into two parts, each defined by the comparison of the chord length with $d$. A hyperspherical cap, by default, 
is a hypersphere part cut by a hyperplane, hence, the latter parts are hyperspherical caps, i.e.

\begin{figure}[htb]
\centering
\includegraphics[width=0.5\textwidth]{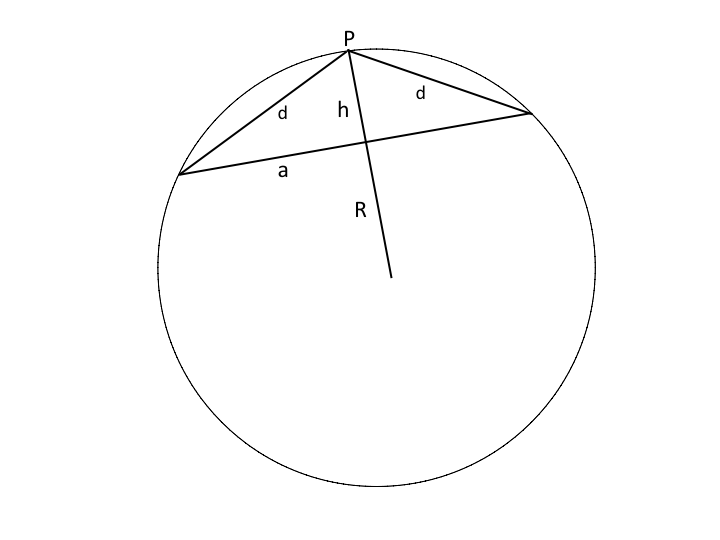}
\caption {A hyperspherical cap and the relation of the maximum distance $d$ from a point $P$, the hyperspherical cap height $h$ and its radius $a$.}
\label{fig:chord_circle}
\end{figure}

\begin{proposition}
The locus of the $N$-sphere points that have distance $D$, $D \leq d$ from a point on it is a hyperspherical cap of radius 
$a=\sqrt{d^2-\frac{d^4}{4R^2}}$. 
\label{theorem1}
\end{proposition}
Proposition \ref{theorem1} implies that the cdf $F_N(d)$ is given as the ratio of a hyperspherical cap surface to the hypersphere surface. Before estimating $F_N(d)$ 
it is reminded that for each N-sphere point $p'= \{p'_{i1}, ~p'_{i2}, ~p'_{i3},~...~, p'_{iN}\}$ with $d(p,p') \leq d$ there is a point 
$p''= \{-p'_{i1},~-p'_{i2},~-p'_{i3},~...~, -p'_{iN}\}$ for which $d(p,p'') \geq \sqrt{4R^2-d^2}$, and vice versa. As a result:

\begin{equation}
\label{eq:symmetry_sphere} F_N(\sqrt{4R^2-d^2}) = 1-F_N(d), d \leq \sqrt{2}R
\end{equation}
Due to Eq. (\ref{eq:symmetry_sphere}), only $F_N(d)$ for $d \leq \sqrt{2}R$ (i.e. corresponding to hyperspherical caps less or equal than a hemi-hypersphere) is 
required. This part of the cdf is estimated using the surface $A^{cap}_N(R)$ of a hyperspherical cap that is smaller than a hyper-hemisphere \cite{SLi11}:

\begin{equation}
\label{eq:li_surface} A^{cap}_N(R) = \frac{1}{2}A_N(R)I_{sin^2\phi}(\frac{N-1}{2},\frac{1}{2})
\end{equation}
In Eq. (\ref{eq:li_surface}), $N$ is the hypersphere dimension, $R$ its radius, $A_N(R)$ the hypersphere surface, $\phi$ the colatitude angle \cite{SLi11} and $I$ the regularised incomplete beta function \cite{EWeisstein4} given by

\begin{equation}
\label{eq:incomplete_beta} I_x(a,b) = \frac{B(x;a,b)}{B(a,b)} = \frac{\int_0^x t^{a-1} (1-t)^{b-1}dt}{\int_0^1 t^{a-1} (1-t)^{b-1}dt}
\end{equation}
In order to eliminate the colatitude angle from Eq. (\ref{eq:li_surface}), we use the fact that $h = (1-cos\phi)R$, where $h$ is the cap height. Since the maximum 
distance $d$ the cap radius $a$ and the cap height $h$ form a right triangle (Fig. \ref{fig:chord_circle}), the height of the cap is $h = d^2/2R$. Therefore, 
$\phi = cos^{-1} (1-d^2/2R^2)$ and the cdf $F_N(d)$ is as follows:

\begin{proposition}
The cumulative distribution function of the $N$-sphere chord length, $F_N(d)$ is
\begin{equation} \label{eq:hypersphere_cdf} 
\centering
\begin{split}
P(D\leq d) = F_N(d) = \frac{1}{2}I_{\frac{d^2}{R^2}-\frac{d^4}{4R^4}}(\frac{N-1}{2},\frac{1}{2}), d<\sqrt{2}R \\
P(D\leq d) = F_N(d) = 1 - \frac{1}{2}I_{\frac{d^2}{R^2}-\frac{d^4}{4R^4}}(\frac{N-1}{2},\frac{1}{2}), d \geq \sqrt{2}R
\end{split}
\end{equation} \label{prop:hypersphere_cdf}
\end{proposition}

The corresponding pdf $f_N(d)$ is:

\begin{proposition}
The probability density function of the $N$-sphere chord length, $f_N(d)$ is:

\begin{equation} \label{eq:hypersphere_pdf}
f_N(d) = \frac{d}{R^2B(\frac{N-1}{2},\frac{1}{2})}(\frac{d^2}{R^2}-\frac{d^4}{4R^4})^{\frac{N-3}{2}}
\end{equation}
\end{proposition}

\subsubsection{Basic properties of the hypersphere chord length distribution}
\label{sec:spotnvdd}

\begin{figure}[htb]
\begin{center}
\begin{tabular}{cc}
\includegraphics[width=0.2\textwidth]{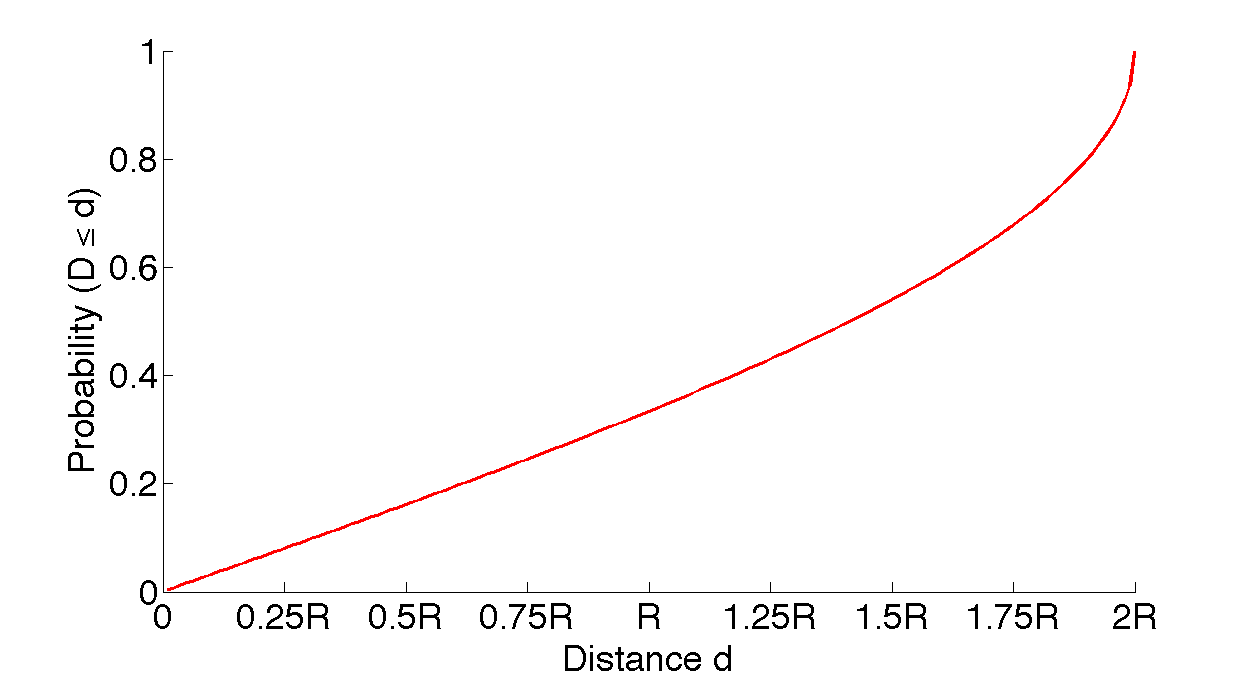} & \includegraphics[width=0.2\textwidth]{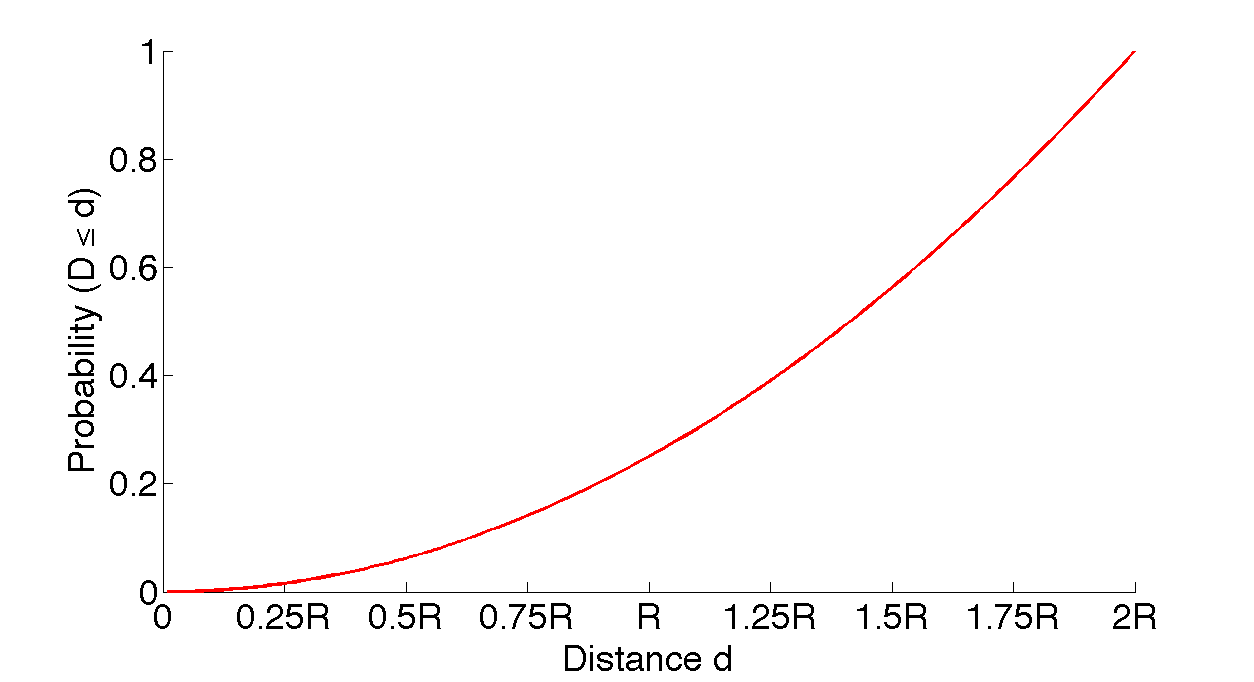} \\
(a) & (b) \\
\includegraphics[width=0.2\textwidth]{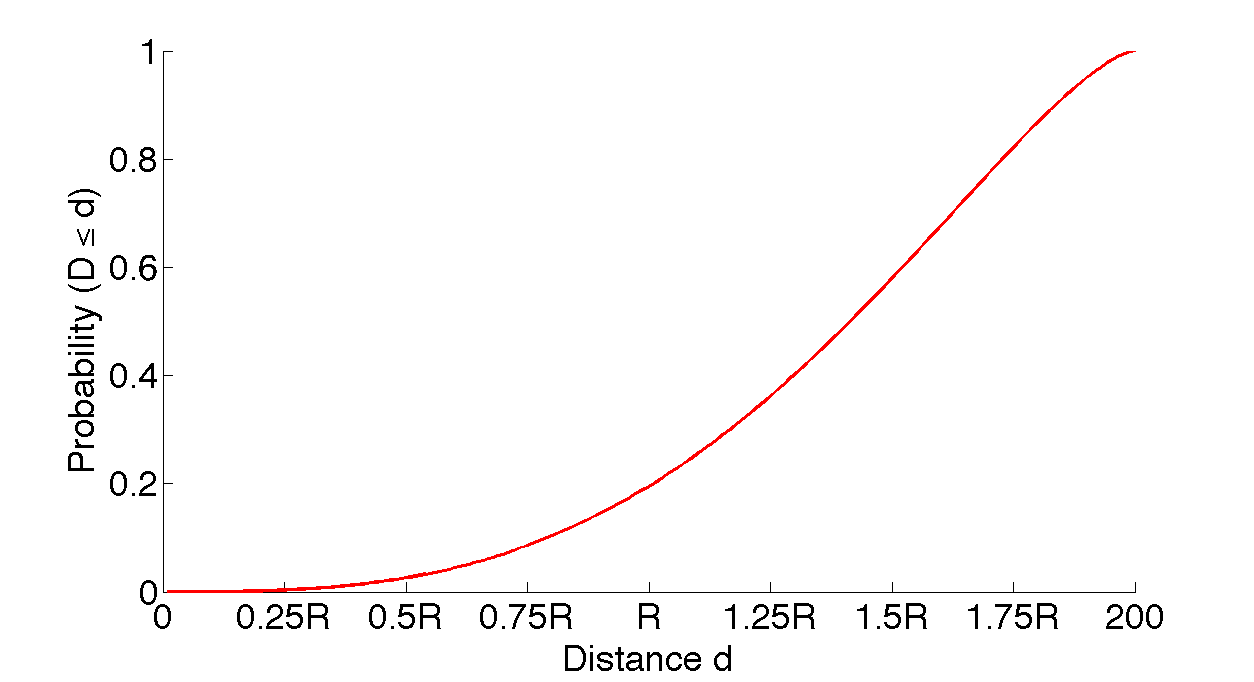} & \includegraphics[width=0.2\textwidth]{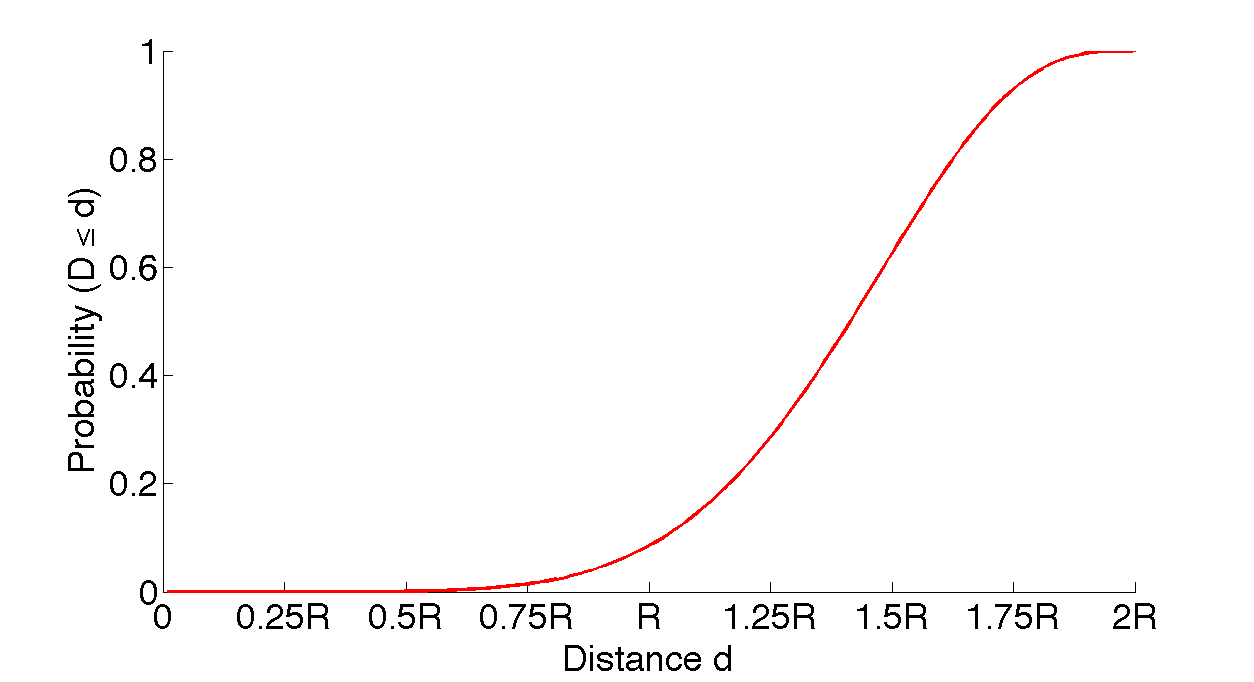} \\
(c) & (d) \\
 \includegraphics[width=0.2\textwidth]{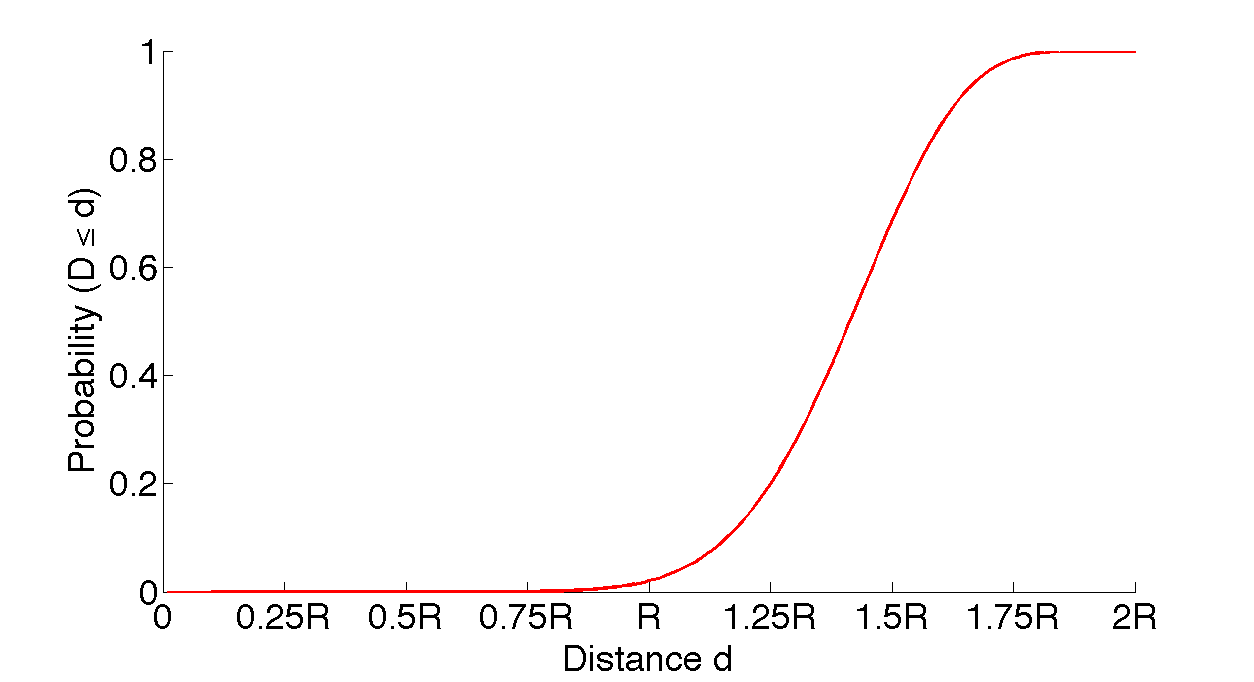} & \includegraphics[width=0.2\textwidth]{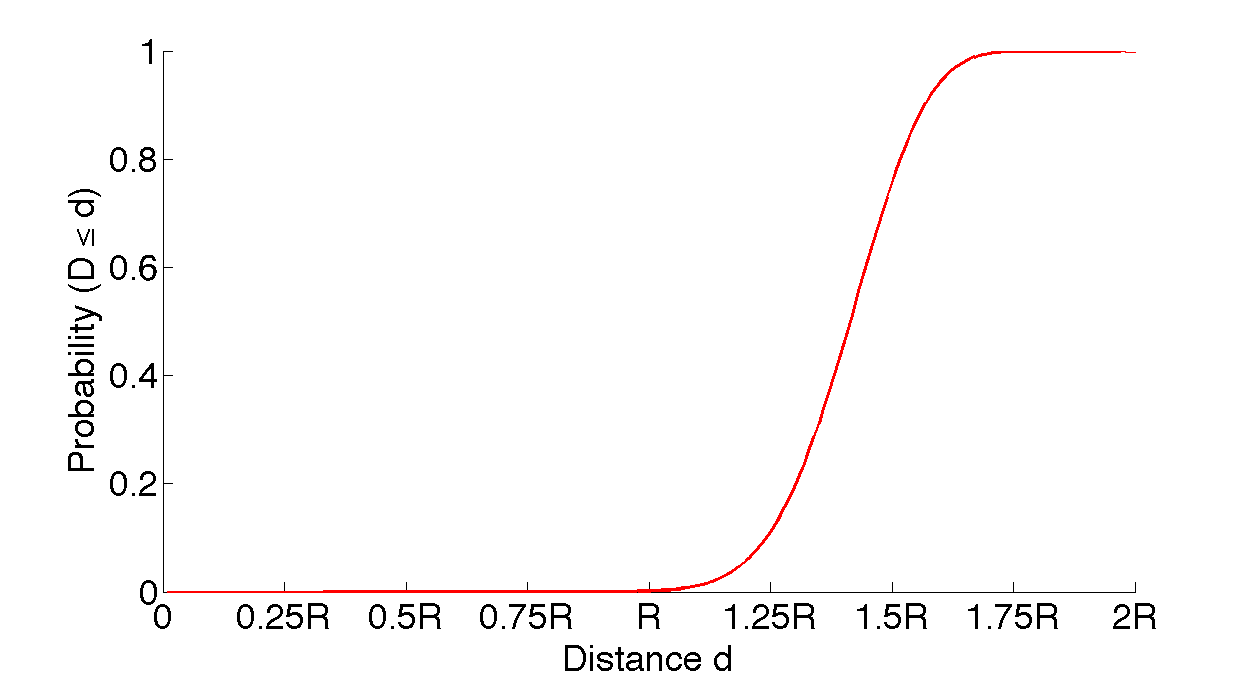} \\
(e) & (f) \\
\end{tabular}
\end{center}
\caption {The cumulative distribution functions. (a) $F_2(d)$ (b) $F_3(d)$ (c) $F_4(d)$ (d) $F_8(d)$ (e) $F_{16}(d)$ (f) $F_{32}(d)$.}
\label{fig:cdf_examples}
\end{figure}

\begin{figure}[htb]
\begin{center}
\begin{tabular}{cc}
\includegraphics[width=0.2\textwidth]{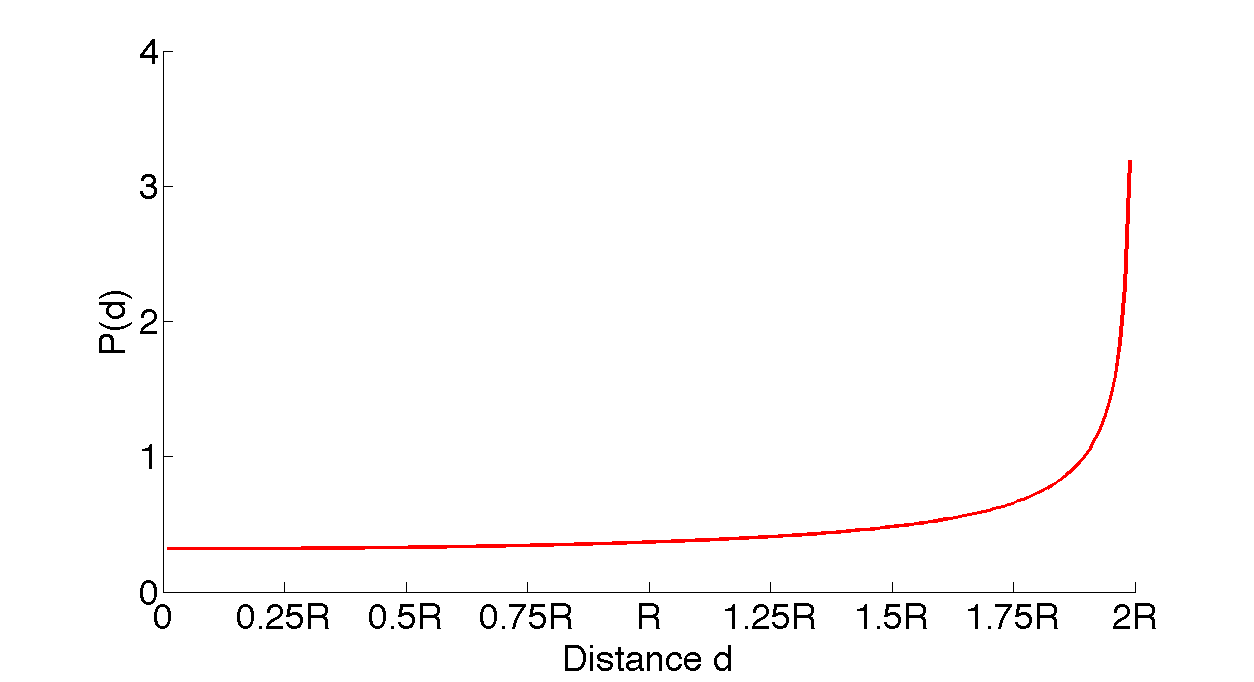} & \includegraphics[width=0.2\textwidth]{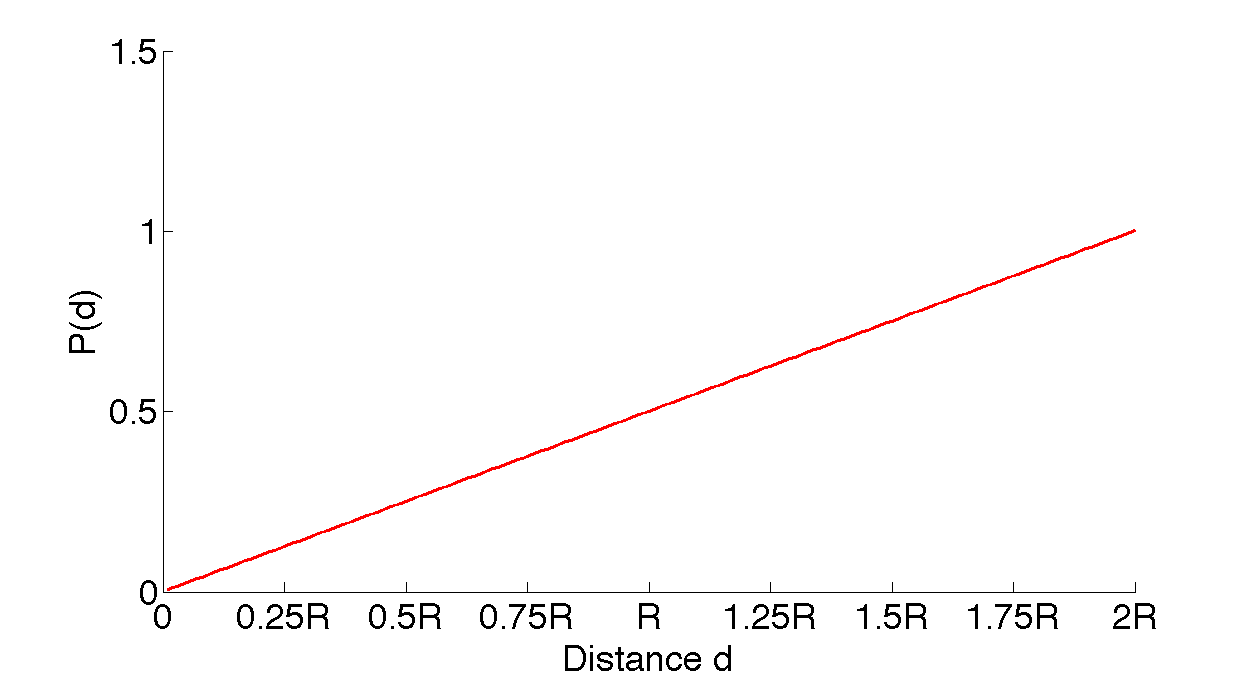} \\
(a) & (b) \\
 \includegraphics[width=0.2\textwidth]{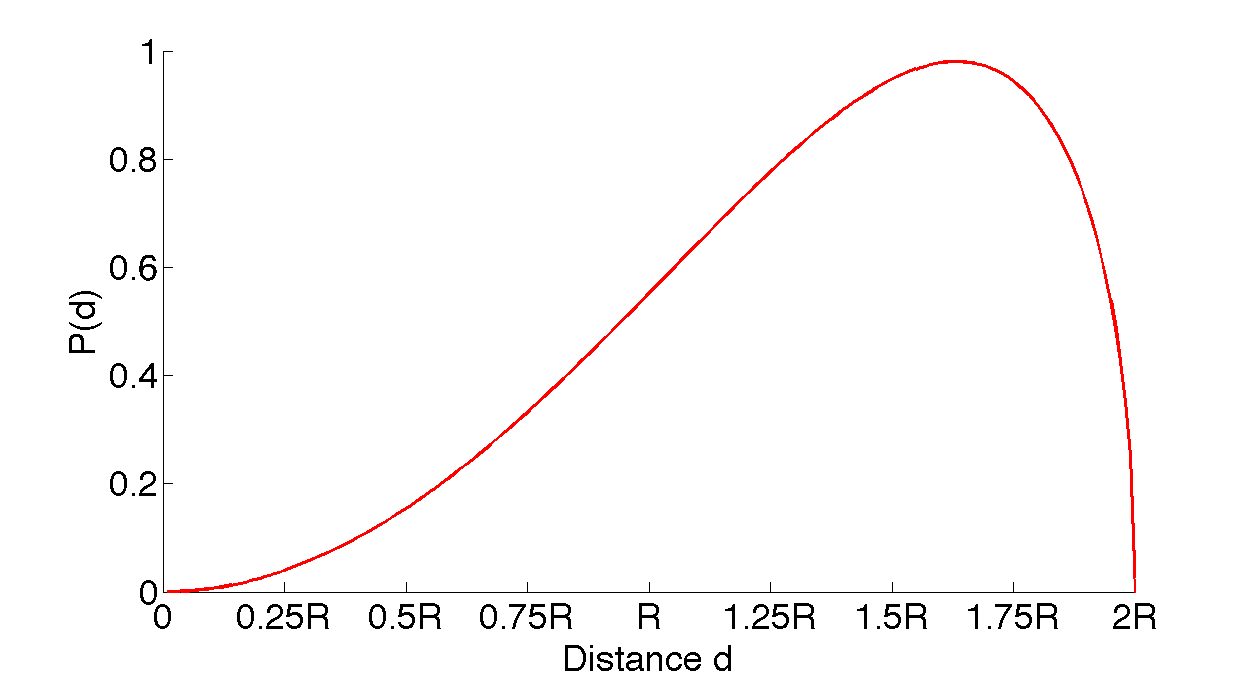} & \includegraphics[width=0.2\textwidth]{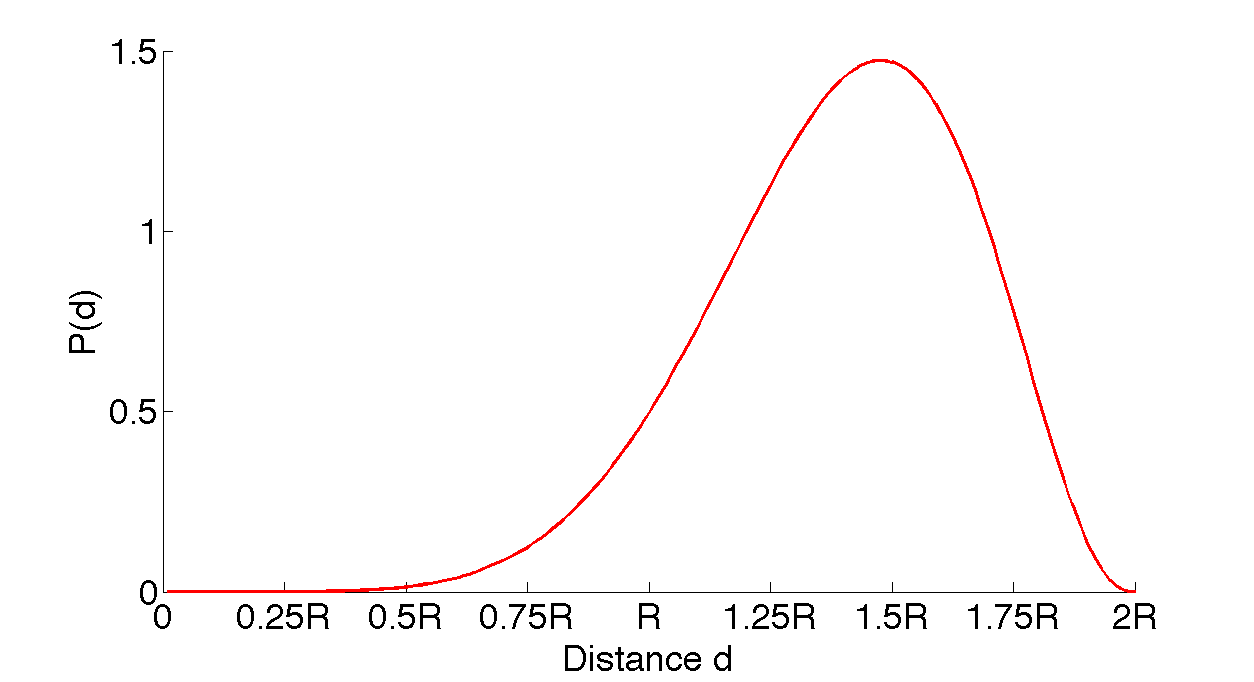} \\
(c) & (d) \\
\includegraphics[width=0.2\textwidth]{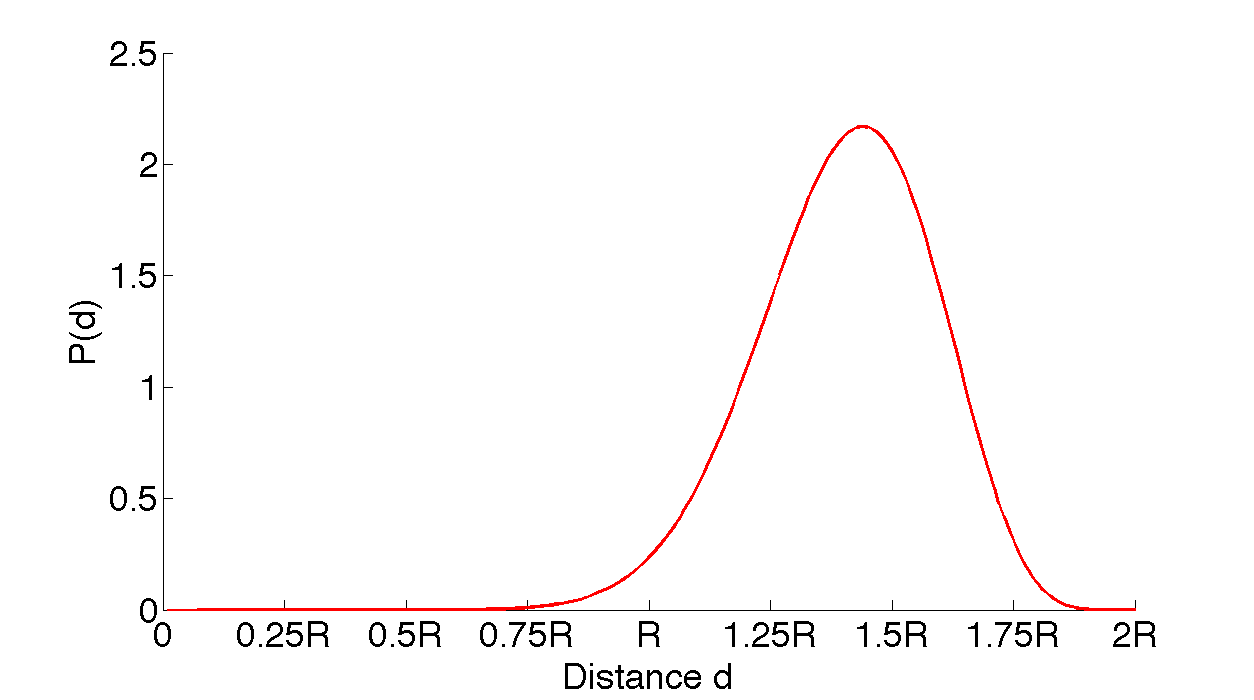} & \includegraphics[width=0.2\textwidth]{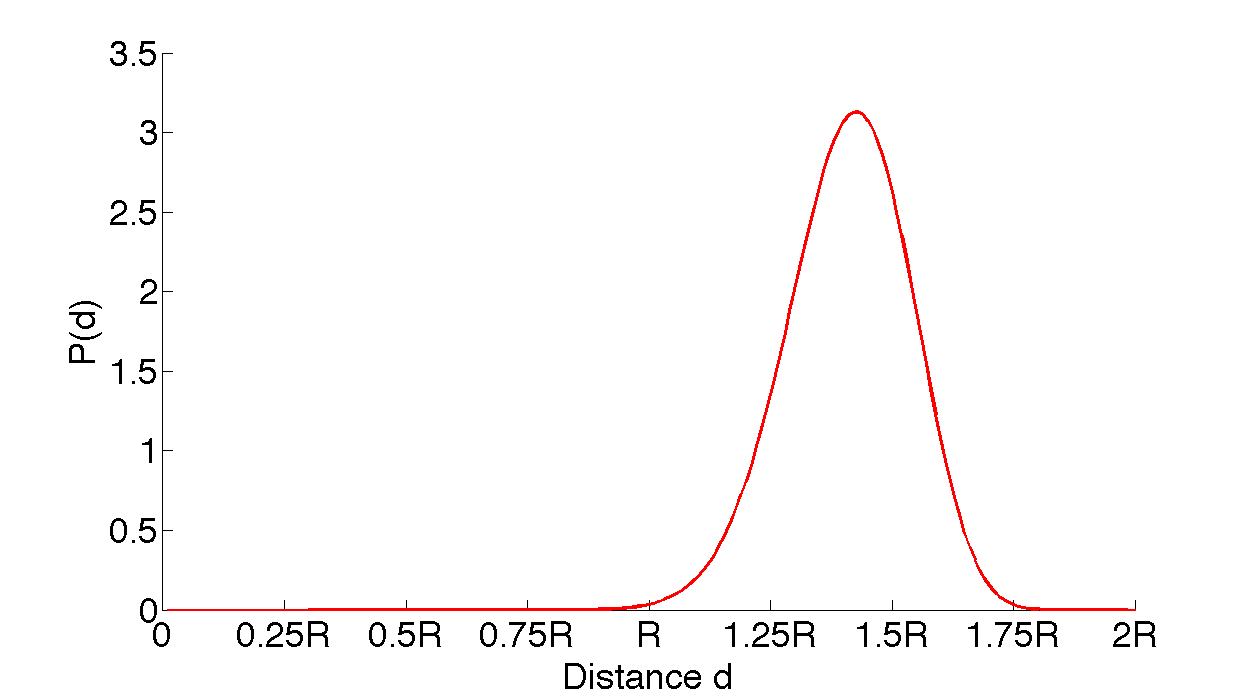} \\
(e) & (f) \\
\end{tabular}
\end{center}
\caption {The probability density functions. (a) $f_2(d)$ (b) $f_3(d)$ (c) $f_4(d)$ (d) $f_8(d)$ (e) $f_{16}(d)$ (f) $f_{32}(d)$.}
\label{fig:pdf_examples}
\end{figure}

Table \ref{tab:basic properties} summarises the chord length distributions of hyperspheres of dimension $2$ to $6$, while the probability density functions and 
cumulative distribution functions for $N = 2, 3, 4, 8, 16, 32$ are shown in Figs. \ref{fig:cdf_examples} and \ref{fig:pdf_examples}, respectively. 

\begin{table*}[htb] \footnotesize
\begin{center}
\begin{tabular}{|c|c|c|c|c|c|}
\hline N & pdf & cdf & Mean & Median & Variance \\ \hline
2 & $\frac{1}{\pi}\frac{1}{\sqrt{1-\frac{d^2}{2R^2}}}$ & $\frac{cos^{-1}(1-\frac{d^2}{2R^2})}{\pi}$ & $\frac{4}{\pi}R$ & $\sqrt{2}$R & $0.379R^2$ \\ \hline
3 & $\frac{d}{2R^2}$ & $\frac{d^2}{4R^2}$ & $\frac{4}{3}R$ & $\sqrt{2}$R & $0.222R^2$ \\ \hline
4 & $\frac{4d^2}{\pi R^3} \sqrt{1-\frac{d^2}{4R^2}}$ & $\frac{cos^{-1}(1-\frac{d^2}{2R^2})}{\pi}-\frac{2}{\pi}(1-\frac{d^2}{2R^2})\sqrt{1-\frac{d^2}{4R^2}}$ & $1.358R$ & $\sqrt{2}$R & $0.156R^2$ \\ \hline
5 & $\frac{3d^3}{4R^4}(1-\frac{d^2}{4R^2})$ & $\frac{3d^4}{16R^4}-\frac{3d^6}{96R^6}$ & $1.371R$ & $\sqrt{2}$R & $0.119R^2$ \\ \hline
6 & $\frac{8d}{3\pi R^2} (\frac{d^2}{R^2}-\frac{d^4}{4R^4})^{3/2}$ & $\frac{2sin^{-1}(\frac{d}{2R})}{\pi} - \frac{\sqrt{4R^2-d^2}(d^7-6R^2d^5+2R^4d^3+12R^6d)}{24\pi R^8} $ & $1.38R$ & $\sqrt{2}$R & $0.0956R^2$ \\ \hline
\end{tabular}
\end{center}
\caption{Basic properties of the N-sphere chord length distribution for $N = 2, 3, 4, 5, 6$.} \label{tab:basic properties}
\end{table*}

The moments about the origin $E(D^k)$ are estimated by using the transform $d^2/4R^2 =u$, which leads to the following equation:

\begin{equation} \label{eq:moments_equation}
E(D^k) = \frac{2^{k+N-2}}{B(\frac{N-1}{2},\frac{1}{2})} B(\frac{k+N-1}{2},\frac{N-1}{2}) R^k
\end{equation}

Hence, for the mean, $\mu$ the following holds:

\begin{proposition}
\label{proposition2} 
The mean value $\mu$ of the $N$-sphere chord length distribution is
\begin{equation} \label{eq:moments_equation_3}
\mu= \frac{\Gamma^2(\frac{N}{2})}{\Gamma(N-\frac{1}{2})\sqrt{\pi}}2^{N-1}R
\end{equation}
\end{proposition}

On the other hand, $E(D^2)$ can be proven to be independent from the hypersphere dimension $N$. Indeed, Eq. (\ref{eq:moments_equation}) for $k=2$ becomes:

\begin{equation} \label{eq:moments_equation_2}
E(D^2) = 2^NR^2 \frac{B(\frac{N+1}{2},\frac{N-1}{2})}{B(\frac{N-1}{2},\frac{1}{2})} = 2^NR^2 \frac{\Gamma(\frac{N}{2})\Gamma(\frac{N+1}{2})}{\Gamma(\frac{1}{2})\Gamma(N)}
\end{equation}
where $\Gamma$ is the Gamma function. Using the following Gamma function property \cite{EWeisstein3}:

\begin{equation} \label{eq:gamma_ratio}
\frac{\Gamma(z)\Gamma(z+\frac{1}{2})}{\Gamma(\frac{1}{2})\Gamma(2z)} = 2^{1-2z}
\end{equation}

and substituting $z=\frac{N}{2}$ in Eq. (\ref{eq:moments_equation_2}) it follows that $E(D^2)=2R^2$. If a point distribution in space is considered as a ''spatial stochastic signal'', then $E(D^2)$ would correspond to the signal power. The independence of $E(D^2)$ from the hypersphere dimension signifies that the ''power'' of the uniform
distribution on a hypersphere is constant in all hyperspheres of equal radius, independently of their dimension.

The variance $\sigma^2$ is straightforwardly estimated by $E(D)$ and $E(D^2)$:

\begin{equation} \label{eq:gamma_ratio}
\begin{split}
\sigma^2 = (2-\frac{\Gamma^4(\frac{N}{2})}{\pi \Gamma^2(N-\frac{1}{2})}2^{2N-2})R^2
\end{split}
\end{equation}

Apart from $E(D^2)$, independent from the dimension is also the median score. By substituting $d=\sqrt{2}R$ to \ref{eq:symmetry_sphere} it follows that 
$F_N(\sqrt{2}R)=0.5, \forall N \geq 2$. $\sqrt{2}R$ is the distance between a ''pole'' and the ''equator'', thus this property is intuitively expected, since it follows by the fact that the two hyper-hemispheres have equal number of points.

Finally, a secondary contribution of the hypersphere distribution is that it allows to estimate the generic solution of the Bertrand problem \cite{APapoulis84}, which refers to the probability $P_R$ of a random chord being larger than the radius $R$. By substituting $d=R$ in Eq. (\ref{eq:hypersphere_cdf}) we get that:

\begin{equation} \label{eq:bertrand}
P_R = P(d\leq R) = \frac{1}{2}I_{\frac{3}{4}}(\frac{N-1}{2},\frac{1}{2})
\end{equation}

$P_R$ is independent from the radius $R$ and rapidly decreasing with respect to the dimension $N$. $P_R$ for $N=2,3,4,5$ is $1/3$, $1/4$, $0.196$ and $0.156$, respectively.


\subsection{Hyper-hemisphere chord length distribution}
\label{sec:hemihyper_distribution}

Apart from the chord length distribution of the whole hypersphere it would be useful to estimate the corresponding distribution of hypersphere sectors, starting with the hyper-hemisphere one.  Without loss of generality it can be assumed that the hyper-hemisphere is the part of the hypersphere for which $p_{iN} \geq 0$. The ''pole'' or, formally speaking, the Chebyshev centre \cite{SBoyd04} of the hyper-hemisphere, i.e. the point that has the minimum maximum distance, is the point $J(0,0,...,0,R)$. The existence of a unique Chebyshev centre (contrary to the hypersphere for which every point has equal maximum distance) implies that points in a hyper-hemisphere are not homogeneous. Therefore, when the number of points $M$ tends to infinity, the rows (and columns) of the distance matrix $d(i,j)$ will not follow the same $f_{NH}(d)$ distribution.

However, the hyper-hemisphere is invariant to rotations around the $p_{N}$ axis, i.e. all points on the surface of the hyper-hemisphere with equal $p_{N}=c$ are produced by the rotation of the point $C(0,0....,0,c',c)$ ($c'^2+c^2=R^2, c' \geq 0$) around $p_{N}$ axis. Since point distance is invariant to rotation, $f_{NH}(d_p)=f_{NH}(d_p')$ if $p_N=p'_N$, where $d_p$ and $d_p'$ is the distance from point $p$ and
$p'$, respectively and $f_{NH}(d_p)$, $f_{NH}(d_p')$ are the respective chord length distributions. As a result, the probability that a hyper-hemispherical chord $D_H$ is smaller than $d$ ($d \leq \sqrt{2}R$) is:

\begin{equation} \label{eq:hemisphere_generic}
P(D_H \leq d) = F_{NH}(d) = \int_0^R P(p_N=c)F_{NH}(d_c) dc
\end{equation}
where $d_c$ is the distance from the point $C(0,0,...,0,c',c)$. A point in the hyper-hemisphere has $p_N \geq c$ if-f it belongs on a hyperspherical cap centered in the
hyper-hemisphere pole with colatitude angle $\phi = cos^{-1}((C\cdot J)/R^2)=cos^{-1}(c/R)$. By equation \ref{eq:li_surface} it follows that:

\begin{equation} \label{eq:firstpointhemi}
P(p_N \geq c) = I_{sin^2\phi}(\frac{N-1}{2},\frac{1}{2}) = I_{1-c^2/R^2}(\frac{N-1}{2},\frac{1}{2})
\end{equation}
and, finally, that:

\begin{equation} \label{eq:firstpointhemi2}
P(p_N=c) = \frac{2}{RB((N-1)/2,1/2)}(1-\frac{c^2}{R^2})^{\frac{N-3}{2}}
\end{equation}

On the other hand, a hyper-hemispherical chord with one end in $C$ has a length less or equal than $d$ if-f it belongs in a corresponding hyperspherical cap of centre $C$ and maximum distance $d$. Therefore, $F_{NH}(d_c)=X_N(d,c) \frac{A^{cap}_N(C,d)}{A^H_N}$, where $X_N(d,c)$ is the percentage of the hyperspherical cap of centre $C$ that lies within the hyper-hemisphere, $A^{cap}_N(C,d)$ is the total surface of the hyperspherical cap and $A^H_N$ is the total surface of the hyper-hemisphere. Since $\frac{A^{cap}_N(C,d)}{A^H_N} = I_{\frac{d^2}{R^2}-\frac{d^4}{4R^4}}(\frac{N-1}{2},\frac{1}{2})$, Eq. (\ref{eq:hemisphere_generic}) becomes:

\begin{equation} \label{eq:hemisphere_generic2}
F_{NH}(d) = KI_{\frac{d^2}{R^2}-\frac{d^4}{4R^4}}(\frac{N-1}{2},\frac{1}{2}) \int_0^R  (1-\frac{c^2}{R^2})^{\frac{N-3}{2}} X_N(d,c) dc
\end{equation}
where $K=2/RB((N-1)/2,1/2)$.

Note that the hyperspherical cap of centre $C$, $A^{cap}(C,d)$, is a rotated version of a same-size hyperspherical cap having as a centre the pole $J$, $A^{cap}(J,d)$. The rotation is on the plane that is defined by the centre of the sphere $O$, the pole $J$ and the chord end $C$, i.e. the plane defined by $p_{N-1}$ and $p_{N}$, and the rotation angle is the angle between $OJ$ and $OC$, which in this case is $\phi$. 

$X_N(d,c)$ is determined by the $p_N$ coordinate of $A^{cap}(C,d)$, which is determined by the $p_{N-1}$ and $p_{N}$ coordinates of $A^{cap}(J,d)$. Even though this seems as a $2$-dimensional geometrical problem, it is more complex than that because $p_{N-1}$ and $p_{N}$ are correlated with the rest of the coordinates through the hypersphere equation. Still, $X_N(d,c)$ is the percentage of $A^{cap}(J,d)$ points for which $-sin(\phi)p_{i(N-1)}+cos(\phi)p_{i(N)} \geq 0$.

A first remark is that if $p_{i(N-1)} \leq 0$ then $-sin(\phi)p_{i(N-1)}+cos(\phi)p_{i(N)} \geq 0$, because $0 \leq \phi \leq \pi/2$ and $p_{i(N)} \geq 0$. The inequality $p_{i(N-1)} \leq 0$ holds for half of $A^{cap}(J,d)$ points because the (N-1)-coordinate of the pole $J$ is $0$ and $OJ$ is an axis of symmetry of $A^{cap}(J,d)$. Therefore, $X_N(d,c) \geq 1/2$. Moreover,  the integral of $P(p_N=c)$ is $1$ because $P(p_N=c)$ is a pdf. By substitution to Eq. (\ref{eq:hemisphere_generic2}) we confirm the following intuitive proposition.

\begin{proposition}
\label{propositionhemi1} 
The hyper-hemisphere cdf is larger than the hypersphere cdf $\forall d \leq \sqrt{2}R$, i.e.
$F_{NH}(d) \geq F_N(d) = \frac{1}{2} I_{\frac{d^2}{R^2}-\frac{d^4}{4R^4}}, ~ \forall d \leq \sqrt{2}R$ 
\end{proposition}
As a matter of fact, $X_N(d,c)$ equals to $1$ if the rotation angle is sufficiently small. To estimate the range of $c$ for which $X_N(d,c)=1$ it is reminded that the part of $A^{cap}(C,d)$ that lies within the hyper-hemisphere is the cut of the hypersphere with two hyperplanes, $L: p_{N}=0$ and $L': \frac{c'}{R} p_{N-1}+\frac{c}{R} p_{N} = R(1-\frac{d^2}{2R^2})$. The cap $A^{cap}(C,d)$ lies entirely within the hyper-hemisphere (i.e. $X_N(d,c)=1$) if-f the hyperplane intersection happens outside the hypersphere. This implies that:

\begin{equation} \label{eq:hemispherexn}
X_N(d,c)=1 \iff d\sqrt{1-\frac{d^2}{4R^2}} \leq c \leq R 
\end{equation}
The integral $ \int_{d\sqrt{1-\frac{d^2}{4R^2}}}^R  (1-\frac{c^2}{R^2})^{\frac{N-3}{2}} dc$, by substituting $c^2/R^2=t$, becomes $\frac{R}{2}\int_{\frac{d^2}{R^2}-\frac{d^4}{4R^4}}^1 (1-t)^{(N-3)/2}t^{-1/2}dt$. Therefore, 

\begin{equation} \label{eq:hemisphere_generic3}
F_{NH}(d) = I_{\frac{d^2}{R^2}-\frac{d^4}{4R^4}}(\frac{N-1}{2},\frac{1}{2}) (1-I_{\frac{d^2}{R^2}-\frac{d^4}{4R^4}}(\frac{1}{2},\frac{N-1}{2})) + I_1
\end{equation}
where $I_1$ is Eq. (\ref{eq:hemisphere_generic2}) with the upper integral limit changed according to Eq. (\ref{eq:hemispherexn}) to $d\sqrt{1-\frac{d^2}{4R^2}}$.

\begin{figure}[htb]
\centering
\includegraphics[width=0.52\textwidth]{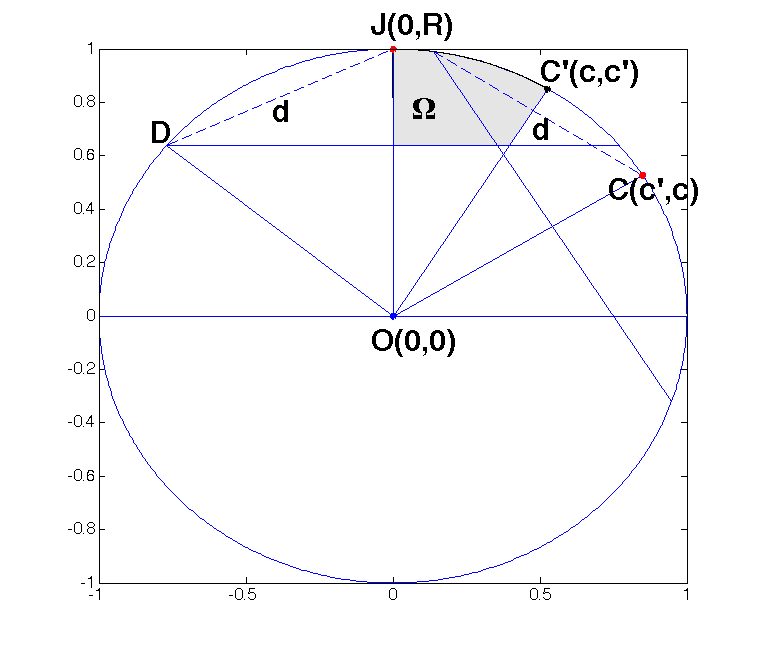}
\caption {$X_N(d,c)$ as the ratio of the shaded area $\Omega$ in relation to the hyperspherical cap with maximum distance $d$.}
\label{fig:semicircle_fig}
\end{figure}

The determination of $X_N(d,c)$ in the case that $X_N(d,c)<1$ (Fig. \ref{fig:semicircle_fig} is a rather challenging problem, which however can be linked to a single surface ratio:

\begin{equation} \label{eq:hemispherexn2}
X_N(d,c)=\frac{1}{2} + \frac{A_\Omega}{A^{cap}(J,d)}
\end{equation}
where $A_\Omega$ is the area of the locus for which $-sin(\phi)p_{i(N-1)}+cos(\phi)p_{i(N)} \geq 0$, $p_{i(N-1)} \geq 0$ (Fig. \ref{fig:semicircle_fig}). 

The area of $A_\Omega$ can be estimates using the intersection of two hyperspherical caps, which has been recently examined in detail \cite{YLee14}. Using the taxonomy of \cite{YLee14}, this corresponds to case No. $9$, i.e. with axis angle $\theta_v$ less than $\pi /2$, and the two hyperspherical caps angles $\theta_1 \in [0~ \pi/2)$ and $\theta_2=\pi /2$. According to \cite{YLee14}, the hyperspherical cap part $X'$ that does not intersect with the hyper-hemisphere $X'=A^{cap}(J,d)/2 - A_\Omega$  is as follows:

\begin{equation} \label{eq:hemispherexnnon}
X' = \frac{\pi ^{\frac{N-1}{2}}}{\Gamma(\frac{N-1}{2})}R^{N-1}\int_{l_1}^{l_2} sin\phi ^{N-2}I_{1-c^2}((N-1)/2,1/2)d\phi
\end{equation}
where $l_1=sin^{-1}(c/R)$ and $l_2=cos^{-1}(1-d^2/2R^2)$. Estimating $X'$ through (Eq. \ref{eq:hemispherexnnon}) and replacing to (Eq. \ref{eq:hemisphere_generic3}) gives the generic formula of hyper-hemisphere chord length distribution. This is a rather challenging task and leads to complex and lengthy expressions even for small values of $N$. As an example, the hyper-hemisphere chord length cdf for $N=4$ is given:

\begin{proposition}
\label{propositionhemi1} 
If $N=4$, the probability that a hyper-hemisphere chord is less or equal than $d$, $d \leq \sqrt{2}R$ $F_{NH}(d)$ is $F_{NH}(d) = P_1(d) - P_2(d) + P_3(d)$, where:
\begin{equation} \label{eq:hemispherexnp1}
P_1(d) =  I_{\frac{d^2}{R^2}-\frac{d^4}{4R^4}}(\frac{N-1}{2},\frac{1}{2}) (1-\frac{1}{2}I_{\frac{d^2}{R^2}-\frac{d^4}{4R^4}}(\frac{1}{2},\frac{N-1}{2}))
\end{equation}
\begin{equation} \label{eq:hemispherexnp2}
P_2(d) =  \frac{2I_{\frac{d^2}{R^2}-\frac{d^4}{4R^4}}(\frac{N-1}{2},\frac{1}{2})[(1-\frac{d^2}{2R^2})^2-(1-\frac{d^2}{2R^2})^N]}{(N-2)\pi B(\frac{N-1}{2},\frac{1}{2})I _{\frac{d^2}{R^2}-\frac{d^4}{4R^4}}(\frac{3}{2},\frac{1}{2})}
\end{equation}
\begin{equation} \label{eq:hemispherexnp3}
P_3(d) =  \frac{2I_{\frac{d^2}{R^2}-\frac{d^4}{4R^4}}(\frac{N-1}{2},\frac{1}{2})\int_0^{asin(\sqrt{\frac{d^2}{R^2} - \frac{d^4}{4R^4}})} \theta cos\theta ^{N-2} d\theta}{\pi B(\frac{N-1}{2},\frac{1}{2})I _{\frac{d^2}{R^2}-\frac{d^4}{4R^4}}(\frac{3}{2},\frac{1}{2})}
\end{equation}
\end{proposition}
Analogous closed-form expressions of the hyper-hemisphere chord length distribution can be estimated using (Eq. \ref{eq:hemispherexnnon}) and (Eq. \ref{eq:hemisphere_generic3}) if required. 

The cdf estimation is completed for $d \geq \sqrt{2}R$ by revisiting the hyperspherical symmetry of Eq. (\ref{eq:symmetry_sphere}) and taking into account that $p'$ and $p''$ belong to different hyper-hemisphere. Therefore, for each pair of points that belong to the same hyper-hemisphere and have a distance $d$ there is a pair of points that belong to different hyper-hemispheres and have a distance $d' = \sqrt{4R^2-d^2}$ and vice versa. This property defines an equation between the cdf of a hyper-sphere and the cdf of a hyper-hemisphere, which leads to the following property for the hyper-hemispherical cdf for $d \geq \sqrt{2}R$:

\begin{proposition}
\label{propositionhemi2} 
The probability that a hyper-hemisphere chord is less or equal than $d$, $d \geq \sqrt{2}R$ $F_{NH}(d)$ is $F_{NH}(d) = 2F_N(d)+F_{NH}(\sqrt{4R^2-d^2})-1$, where $F_N$ and $F_{NH}$ are the hyper-spherical and hyper-hemispherical cdfs for $d\sqrt(2)R$, respectively.
\end{proposition}

\subsubsection{Basic properties of the hyper-hemisphere chord length distribution}
\label{sec:basice_properties_hyperhemi}

Following the above analysis one can estimate both the cdf and the pdf of the hyper-hemisphere chord length distribution for any $N$. However, it is apparent that the compactness of the (whole-)hypersphere case, i.e. the hyperspherical chord length distribution, is lost, thus implying that estimating analytical formulas of the chord length distribution of widely used hyperspherical segments and/or sectors is expected to be a rather challenging task.

Another difference between the hypersphere and hyper-hemisphere distribution is the fact that for $d=\sqrt{2}R$ the cdf is no longer independent from the hypersphere dimension (let alone equal to $0.5$). For example, by substituting $d=\sqrt{2}R$ in the equations of proposition \ref{propositionhemi1} it follows that $P_1(\sqrt{2}R)=1/2$, $P_2(\sqrt{2}R) =0$, i.e. $F_{NH}(\sqrt{2}R) - F_{N}(\sqrt{2}R) = P_3(\sqrt{2}R)$. In this case, the divergence between the hyper-hemisphere and the hypersphere cdf value for $d=\sqrt{2}R$, $P_3(\sqrt{2}R)$, is:

\begin{equation} \label{eq:hemisphere_diversqr2}
P_3(\sqrt{2}R) = \frac{2 \int_0^{\frac{\pi}{2}} \theta cos\theta ^{N-2} d\theta}{\pi B(\frac{N-1}{2},\frac{1}{2})} \overset{(N=4)} = 1/4-1/\pi ^2
\end{equation}
Finally, the independence of the second moment, $E(D^2)$, from the dimension $N$ does not hold for the hyper-hemisphere. However, the gradual decrease of the variance with dimension is also apparent in the hyper-hemisphere case, as shown in Table \ref{tab:basic_properties_hemi}. Table \ref{tab:basic_properties_hemi} summarises the basic properties of all hyper-hemisphere chord length distributions for dimensions $3$ to $6$

\begin{table}[htb] \footnotesize
\begin{center}
\begin{tabular}{|c|c|c|c|}
\hline N & Mean & Median & Variance \\ \hline
3 & $1.124R$ & $1.147R$ & $0.217R^2$ \\ \hline
4 & $1.218R$ & $1.249R$ & $0.157R^2$ \\ \hline
5 & $1.268R$ & $1.296R$ & $0.121R^2$ \\ \hline
6 & $1.299R$ & $1.322R$ & $0.0985R^2$ \\ \hline
\end{tabular}
\end{center}
\caption{Basic properties of the hyper-hemisphere chord length distribution for dimension 3 to 6.} \label{tab:basic_properties_hemi}
\end{table}

\section{Hypersphere chord length distribution as a uniformity measure}
\label{sec:point_uniform}

As already mentioned, some of the most interesting properties of the hypersphere chord length distribution arise from the fact that this is the limit distribution of the distances of uniformly selected hypersphere points. To summarise, the hypersphere chord length distribution is the limit distribution of $3$ (related but distinct) distributions:

\begin{enumerate}
\item The distance distribution of $M$ point-pairs $||p_i-p'_i||, i=1,2,...,M$, if the $2M$ relevant points $p_i$ and $p'_i$ are independently selected from a uniform random distribution.
\item The intra-distance distribution of a set of $M$ points $p_i, i=1,2,...,M$, if the $M$ relevant points are independently selected from a uniform random distribution, before the $M(M-1)/2$ pairwise distances $||p_i-p_j||, i,j=1,2,...,M, i \neq j$
are estimated.
\item The distance distribution of a set of $M-1$ points $p_i, i=1,2,...,M-1$ from a fixed point $p_0$ if both $p_i$ and $p_0$ are selected from a uniform random distribution before the $M-1$ pairwise distances $||p_i-p_0||, i=1,2,...,M-1$ are estimated. \end{enumerate}
The second and the third distribution allow the hypersphere chord length distribution to be used as an uniformity measure, as is described in the current section.

More specifically, in order to quantify the ''uniformity'' of an input point distribution on a N-sphere, the L1 distance is used. If the intra-distance distribution of the input point distribution is $g_N$ then the $L_1(g)$ uniformity measure is defined as follows:

\begin{equation} \label{eq:hypesphere_uniform_distance}
L_1(g) = \int_0^2 |g_N(x)-f_N(x)| ~dx 
\end{equation}
where, $f_N$ is the hypersphere chord length distribution. Note that this uniformity measure can be used to quantify the uniformity of all $3$ types of point distance distributions that are described above. 

The reason for selecting $L_1$ is double; firstly, it satisfies the metric conditions, thus defining a metric space; secondly, it was experimentally found that $L_1(g)$ convergence rate to $0$ for a uniform distribution is $k^{-1/2}$, where $k$ is the number of point-pairs that are included in the $g_N$ distribution ($k=M$, $k=M(M-1)/2$ and $k=M-1$, respectively, for the $3$ examined types of point distance distributions). This allows the experimental computation of ''confidence intervals'' for $L_1(g)$ even when $M$ takes an impractically large value. Initially, uniform pointsets of size $M'$ ($M' \ll M$) are generated on the N-sphere and $L_1$ values are sorted, before acquiring the $\alpha\%$-largest $L_1$ value and finally extrapolating for pointsets of size $M$. This value is the threshold with which the input distance distribution $L_1(g)$ is compared to determine whether it is uniform or not.

Elaborating on this idea, based on the computational cost of iteratively estimating pairwise distances, $L_1$ can be used to qualitatively assess whether an $N$-dimensional point sample $S$ (consisting of $M$ points, and having an intra-distribution $g$) originates from a uniform N-sphere (or N-hemisphere) distribution following on of the three following approaches:

\begin{itemize}
\item If $S$ dimension $N$ and sample size $M$ imply a non-prohibitive computational cost, then $Q$ uniformly distributed point sets of size $M$ and dimension $N$ are randomly generated and $L_1$ is estimated for all of them (as well as for $S$). If the $\alpha\%$-largest $L_1$ value of $Q$ is smaller than $L_1(g)$ then $S$ can be declared as non-uniform with confidence $(100-\alpha)\%$.
\item If $S$ dimension $N$ and sample size $M$ imply a prohibitive computational cost for estimating $L_1$ for $Q$ uniformly distributed point sets ($Q \gg 1$) but not for estimating $L_1(g)$, then the difference with the previous case is that the $\alpha\%$-largest $L_1$ value of $Q$ is estimated using sets of $M'$ $N$-dimensional points ($M' \ll M$) and extrapolating for $M$.
\item If the computational cost needs to be further reduced then a point of $S$ is fixed and the distributions of the distances from this point are estimated. These distributions are still expected to have as a limit distribution the hypersphere chord length distribution, while the associated computational complexity is linear (instead of quadratic).
\end{itemize}

The above tests are designed to identify non-uniform spatial distribution, thus can not securely confirm uniformity. In practice, this is rarely expected to be of major importance because the uniformity-measurement objective is usually to assess whether the points span the hypersphere in a way that is compatible with the uniform distribution; not to mathematically confirm that they actually originate from a ''pure'' uniform hypersphere distribution. For example, if an algorithm has a large number of hyperparameters and its exhaustive evaluation in the hyperparameter space is computationally expensive, a straightforward approach would be to sub-sample the hyperparameter space ''uniformly'', selecting a small number of points (i.e. hyperparameter combinations). In this case, uniformity of the set of hyperparameter-points is not a strict theoretic requirement but a rather loose condition so as to ensure that the evaluation does not omit a large neighbourhood in the hyperparameter space. The above tests would be sufficient to assess whether the hyperparameter-points were ''uniformly'' selected or not.

Apart from the qualitative evaluation, $L_1$ can be used to generate a quantitative uniformity measure, specifically, the size of the maximum uniform subset $M_u, M_u \leq M$ of a N-sphere pointset $S$ ($|S|=M$). The relevant presentation starts by reminding that a pointset $S$ can be considered as a mixture of a uniform subset $S_u$ ($|S_u|=M_u$) and a non-uniform subset $S_c$ ($|S_c|=M_c = M-M_u$). The intra-distance distribution $g_N(S)$ is a weighted average of $3$ distance distributions: (a) the (intra-)distance distribution $g_u$ of pairs selected from $S_u$ (b) the (inter-)distance distribution $g_{uc}$ of pairs in which one point is selected from $S_u$ and one point is selected from $S_c$ and (c) the (intra-)distance distribution $g_c$ of pairs selected from $S_c$, the respective weights being $W_a = M_u(M_u-1)/2k$, $W_b = M_uM_c/k$ and $W_c = M_c(M_c-1)/2k$, where $k=M(M-1)/2$ (note that $W_a+W_b+W_c=1$).

Since $S_u$ are selected from a uniform distribution, $L_1(g_u)=0$ (it is assumed that $M,M_u,M_c = \infty$). Regarding the inter-distance distribution $L_1(g_{uc}$, this is also $0$ because $g_{uc}$ is a sum of $S_c$ distance distributions in which one point of the pair is fixed on the hypersphere while the second is uniformly selected, i.e. $S_c$  (identical) hypersphere chord length distributions (as implied by the third distribution for which the hypersphere chord length distribution is the limit distribution). Therefore, $L_1(g) = W_cL_1(g_c)$.

Since $L_1(g) \equiv L$ can be straightforwardly estimated by the pointset $S$, and $L_1(g_c) \leq 2$ (this follows by the inequality $|g_N(x)-f_N(x)| \leq 1, \forall x$), then a lower limit for $W_c$ is $L/2$. Since $M_c$ and $M$ are infinite, then $W_c = (M_c/M)^2$, i.e. $M_c/M \geq \sqrt{L/2}$. The ratio $M_c/M$ is the percentage of non-uniform points in $S$, therefore, what has been proven is the following:

\begin{proposition}
If a set of points $S$ on the N-sphere have an intra-distance distribution for which the $L_1$ distance from the N-sphere chord length distribution is $L$, then the maximum percentage of uniform points in $S$ is equal to $1-\sqrt{L/2}$.
\label{maximumunfrmsubs} 
\end{proposition}
Proposition \ref{maximumunfrmsubs} constraints the maximum size $M_u$ of a uniform subset of a pointset defined on a hypersphere. In practical applications $S$ is finite, hence there is an uncertainty in the $M_u$ upper limit because $g_u$ and $g_{uc}$ have not fully converged yet to the hypersphere chord length distribution, and $W_c$ is only approximately equal to $(M_c/M)^2$. In this case, the implied maximum size $M_u$ should be checked so as to ensure that the uncertainty does not significantly tamper the upper limit. 

For example, if $M=5,000$, $L=0.5$ and $N=3$ then Proposition \ref{maximumunfrmsubs} gives $M_u=2,500$. By instantiating $1,000,000$ uniform pointsets of dimension $3$ and size $2,500$ it is estimated that the average divergence from the (3-)sphere chord length distribution is $0.0042$ and the $1\%$-largest divergence $0.0052$. Since $W_a \approx 1/4$, $L_1(g_u) \leq 0.0015$, i.e. negligible in comparison to $L$. On the other hand, $L_1(g_{uc})$ is theoretically more difficult to eliminate because it depends on the averaging of $2,500$ uniform sample distributions, each one generated by $2,500$ distance samples, which is not straightforward to theoretically analyse because the distributions are not mutually independent. However, due to the fact that averaging exhibits a powerful noise reduction effect, it was experimentally found that usually $L_1(g_{uc})  \approx L_1(g_u)$. Since both $L_1(g_{uc})$ and $L_1(g_{u})$ are negligible, $M_u$ estimation can be considered accurate enough.

A more safe estimation of $L_1(g_{uc})$ is a sideproduct of a novel algorithm that proceeds to estimate the actual maximum uniform subset $S_u$ (Algorithm \ref{alg:uniformsubsetestim}. This algorithm is a Monte-Carlo voting scheme which generates uniform sets of size $M_u$ (independent to $S$) and project them to $S$. In this algorithm, $M_u$ value controls the size of the generated uniform sets. This kind of an algorithm, which separated uniform for the non-uniform subset of a set can be very useful in several applications. For example, in clustering applications, points that can be generated from an uniform distribution may be assumed to not belong to any (locally defined) class, hence identifying and discarding the maximum uniform subset will typically disambiguate the inter-classes boundaries. 

\begin{algorithm}\small
\begin{algorithmic}
\item[Input:] A set $S$ of $M$ $N$-dimensional points defined on the N-sphere, with a distance distribution $g$, number of repetitions $E_m$, current repeat $E=1$, VoteVector(i)=0, $i=1,2,...M$.
\item[1: ] Estimate $L_1(g)$ using Eq. (\ref{eq:hypesphere_uniform_distance}).
\item[2: ] Estimate $M_u$ using proposition \ref{maximumunfrmsubs}.
\item[3: ] Randomly select a uniform N-dimensional set $S_E$ ($|S_E|=M_u$).
\item[4: ] For all points in $S_E$ estimate their nearest neighbour in $S$, i, and assign VoteVector(i) = VoteVector(i)+1.
\item[5: ] If $E \geq E_m$ return the $M_u$ points with the largest VoteVector corresponding values, else $E=E+1$ and go to Step 3.
\end{algorithmic}
\caption{Algorithm for the estimation of the maximum-size uniform subset of a pointset defined on a hypersphere.} \label{alg:uniformsubsetestim}
\end{algorithm}.
Finally, it should be noted that in the case of hyper-hemispherical uniform data only the qualitative analysis conducted in this section stands. As a matter of fact, while the hyper-hemisphere chord length distribution is the limit distribution of a set of $M$ points independently and uniformly selected on a hyper-hemisphere and the convergence rate is still $k^{-1/2}$, proposition \ref{maximumunfrmsubs} does not stand because as explained in Section \ref{sec:hemihyper_distribution} the points in a hyper-hemisphere are not homogeneous. $L_1$ distance is still expected to denote whether a sample originates from a uniform hyper-hemispherical distribution, however, no quantitative conclusions can be derived from the specific $L_1$ value using the techniques presented in this section.

\section{Detecting uniform sets in higher dimensions}
\label{sec:higher_dimensions}

In the previous section we have discussed how the hypersphere chord length distribution can be used to assess the uniformity of a point set defined on a hypersphere. In this section we are discussing whether a uniform subset $S_u$ embedded in a (not necessarily uniform) set $S$ of higher dimension can be identified using the hypersphere chord length distribution. It will be demonstrated that this is practically possible because the distribution of distances from a point that belongs to $S_u$ will be a mixture of two distributions, one of which is the hypersphere chord length distribution.

Let's assume that a uniform subset $S_u$ with dimension $N_u$ and size $M_u$ is embedded into a set $S$ ($S \supset S_u$) of dimension $N$ ($N>N_u$) and size $M$ ($M > M_u$), and $p_i$ a point in $S_u$ then the distribution $g_{iS}$ of the distances from $p_i$ is:

\begin{equation} \label{eq:distance_distribution_embed}
g_{iS}(d) = (M_u/M)f_{N_u}(d) + (1-M_u/M)g'_N(d)
\end{equation}
where $f_{N_u}(d)$ is the $N_u$-sphere chord length distribution and $g'_N(d)$ a generally unknown distance distribution. The $L_1(g)$ distance between $g_{iS}(d)$ and $f_{N_u}(d)$ is

\begin{equation} \label{eq:distance_distribution_l1}
L_1(g) = (1-M_u/M) \int_0^2 |g'_N(x)-f_{N_u}(x)| dx
\end{equation}
On the other hand, if $p_{i'} \notin S_u$ the $L_1$ distance is given by the same formula without the scaling factor $(1-M_u/M)$ and with a generally different $g''_N(x)$ function. If $(1-M_u/M)$ is small enough to cancel the difference between $\int_0^2 |g'_N(x)-f_{N_u}(x)| dx$ and $\int_0^2 |g''_N(x)-f_{N_u}(x)| dx$ then the uniform subset can be identified one point at a time using an information retrieval approach; the $M_u$ smaller $L_1$ distances of $g_{iS}$ ($p_i \in S$) from $f_{N_u}$ would correspond to the $M_u$ points originating from the uniform $N_u$-dimensional distribution.

The aforementioned condition depends on two parameters: (a) the uniform subset relative size $M_u/M$, (b) the ''resemblance'' of the distance distribution of the superset $S$ and the $N_u$-sphere chord length distribution. Regarding the first parameter, Eq. \ref{eq:distance_distribution_l1} confirms the intuitive assumption that the performance is increasing with the uniform-subset relative size. On the other hand, the second parameter is less intuitive and more difficult to decipher.

In general, if the integral $\int_0^2 |g'_N(x)-f_{N_u}(x)| dx$ fluctuates between a value $\mu-\sigma$ and a value $\mu+\sigma$ (for different $p_i \in S$), then the detection of $S_u$ would be facilitated by a large $\mu$ (and a small $\sigma$) value. Therefore, the performance increases when the distance distribution of $S$ is substantially different from the $N_u$-sphere chord length distribution. By examining how the N-sphere chord length distribution is modified with the dimension $N$ we gain more insight about this statement.

By revisiting Proposition \ref{prop:hypersphere_cdf} it can be proven that $f_N(d)$ becomes progressively more narrow and as $N$ approaches infinity $f_N(d)$ approaches $\delta(\sqrt{2}R)$, where $\delta(x)$ is the Dirac delta function \cite{APapoulis84}. Using the Beta function as a trigonometric integral \cite{EWeisstein4}:

\begin{equation} \label{eq:beta_trigonometric}
B(x,y) = 2 \int_0^{\pi/2} (sin\theta )^{2x-1}(cos\theta )^{2y-1} d\theta
\end{equation}
it follows that:

\begin{equation} \label{eq:beta_trigonometric2}
F_N(d) = \frac{\int_0^{sin^{-1}(d^2/R^2-d^4/4R^4)} (sin\theta )^{N-2} d\theta}{\int_0^{\pi/2} (sin\theta )^{N-2} d\theta}, d \leq \sqrt2R
\end{equation}
As $N \rightarrow \infty$, $(sin\theta )^{N-2}$ approaches $0$ for $\theta \neq \pi/2$ and $1$ for $\theta=\pi/2$. Therefore, the nominator of Eq. \ref{eq:beta_trigonometric2} is non-zero if-f $sin^{-1}(d^2/R^2-d^4/4R^4)=\pi/2$, i.e. if-f $d=\sqrt2R$, which means that $F_N(d)=0$ for $d<\sqrt2R$ and $F_N(d)=1$ for $d=\sqrt2R$.

Additionally, the following recursive formula stands for incomplete beta functions \cite{EWeisstein4}:

\begin{equation} \label{eq:incomplete_recursice}
I_x(a+1,b) = I_x(a,b) - \frac{x^a(1-x)^b}{aB(a,b)}
\end{equation}
In the $N$-sphere chord length distribution case $a=(N-1)/2$, $b=1/2$ and $x=d^2/R^2-d^4/4R^4$. Under these constraints (and since $0 \leq x \leq 1$), the second term of the right part of Eq. (\ref{eq:incomplete_recursice}) is always positive in the interval $d \leq \sqrt{2}R$, i.e., the cdf scores that correspond to a fixed $d$ value, $d \leq \sqrt{2}R$ reduce with $N$. The opposite is true in the interval $d \geq \sqrt{2}R$. Hence, as the dimension increases the pdf of Eq. (\ref{eq:hypersphere_pdf}) becomes increasingly more concentrated around $\sqrt{2}R$ (Fig. \ref{fig:pdf_examples}).

Finally, Eq. \ref{eq:incomplete_recursice} can be considered as the equivalent of derivative with respect to the dimension. The ratio between two consecutives ''derivatives'' is as follows:

\begin{equation} \label{eq:second_derivative}
\frac{F_{N+2}(d)-F_{N+1}(d)}{F_{N+1}(d)-F_{N}(d)} = \frac{N}{N+1}(\frac{d^2}{R^2}-\frac{d^4}{4R^4}) \leq 1, d \leq \sqrt2R
\end{equation}
As a result, the $L_1$ distance between two N-sphere chord length distributions for adjacent $N$ values is an decreasing function of the dimension $N$. 

To summarise, $N$-sphere chord length pdf continuously converges to the $\delta (\sqrt2R)$ function. Moreover, for all $d, d \neq \sqrt2R$, $f_N(d)$ is a decreasing and concave function of $N$. Based on these properties, it is possible to determine two special cases that the introduced information retrieval scheme is expected to achieve high detection rate:

\begin{itemize}
\item If $N_u \gg 1$ and the distance distribution of $S$ is not a narrow function around $\sqrt2R$.
\item If $S$ is a uniformly distributed set of dimension $N$ ($N>N_u$) and either $N \gg N_u$ or $(N/N_u) \gg 1$.
\end{itemize}

The discussion conducted in this Section is not valid for hyper-hemispherical uniform subsets. The lack of point homogeneity in hyper-hemispheres prohibits using the distribution of distances from a fixed point as a uniformity descriptor, thus invalidating the introduced information retrieval scheme. However, it can be proven that hyper-hemisphere chord length distribution also converges to the $\delta (\sqrt2R)$ function. This property may be possibly exploited to develop hyper-hemisphere uniform subset detection techniques.

Before finishing the theoretic part of this work, it would be interesting to have a brief discussion about a purely theoretic concept that is rarely examined, the hypersphere of infinite dimensions. The chord length distribution in this case implies that the probability of two points having distance $\sqrt2R$ is $1$. Taking into account that in (3-dimensional) spheres $\sqrt2R$ is the distance between a pole and the equator, if a point in the infinite-dimension hypersphere is arbitrarily selected as a pole, then ''almost all'' (meaning infinitely more than not) other points lie on the equator. Therefore, we reach to the counterintuitive conclusion, that in an infinite dimension space, a sphere and its equator represent (almost) identical concepts.

\section{Application demonstrations and experimental evaluation}
\label{sec:applications}

In this section some examples of the potential use of the hypersphere chord length distribution as a uniformity measure are given. The employed algorithms were designed to be as simple as possible, involving no more than the basic concepts discussed in the theoretic part of this article. The reason for this design principle was double; firstly, the scope of this work was to validate the hypersphere chord length distribution as a uniformity measure that can find a broad range of applications, and not to present an optimised and complex algorithm that was developed to tackle a specific problem; secondly, by keeping the algorithm development in a basic level it is ensured that the achieved performance is induced by the introduced uniformity measure and not by an elaborate algorithm setup.

\subsection{Monitoring uniform-pointset generation algorithms}
\label{subsec:monitoring}

Hypersphere chord length distribution can be used to monitor the generation of pointsets in terms of uniformity, aiming either to optimise the pointset span or to debug the algorithm that has produced them. The first objective mainly refers to the sampling of discrete spaces, in cases that generating an uniform grid is not an option. For example, a desirable property of the initial population in genetic algorithms may be to uniformly span the solution space. In such a case, the initial population can be selected according to its uniformity, measured by the $L_1$ distance of the (projected to a hypersphere) initial population distance distribution from $f_N(d)$, where $N$ is the parameter space dimension.

\begin{figure}[htb]
\centering
\includegraphics[width=0.4\textwidth]{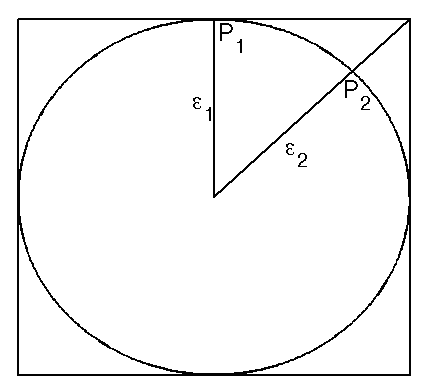}
\caption {An erroneous uniform-point generation technique (in the 2-dimensional space). By selecting randomly and independently $2$ values in the $[-1, 1]$ points inside the square are uniformly defined. Subsequently, they are projected on the circle according to their angle and stored. As a result, all points in line segment $\epsilon_1$ will be projected on $P_1$ while all points in line segment $\epsilon_2$ will be projected on $P_2$. Since $\epsilon_2$ is longer than $\epsilon_1$ the probability of generating $P_2$ is larger than the probability of generating $P_1$, i.e. the final pointset is not uniform.}
\label{fig:error_fig}
\end{figure}

On the other hand, debugging refers to validating algorithms supposed to generate uniform pointsets, especially if simple solutions (such as visual inspection) give ambiguous results. For example, a subtle error in generating uniform pointsets on a $N$-sphere is to initially generate $N$ random values $-1 \leq v_i \leq 1, i=1,2,...N$ and subsequently to normalise the vector $(v_1,v_2,...v_N)$. This approach generates points over the whole hypersphere but not with equal probability (Fig. \ref{fig:error_fig}), i.e. not uniformly.

The non-uniformity of this distribution may be missed due to the fact that it is a mixture of a uniform and a non-uniform distribution, with the uniform component being of large magnitude. For example, in the 2-dimensional case (Fig. 5), the set of points randomly initialised within the circle (before being projected on the circle) constitute the uniform part of the distribution, while the set of points randomly initialised outside the circle (but within the square) the non-uniform. Consequently, $78.54\%$ (approximately equal to $\pi/4$, i.e. the probability of a point being initialised within the circle) of the points follow the uniform circle distribution while $21.46\%$ not. With approximately $4$ out of $5$ points being uniformly distributed, the resulting spatial distribution is difficult to be recognised as non-uniform through graphical means (e.g. a plot of the points). Moreover, the final point distribution is horizontally, vertically and diagonally symmetric, hence binning points in $2$, $4$ or $8$ equal-angle bins would erroneously imply that the distribution is uniform while if more bins are used then it should be ensured that the divergence from uniformity is statistically significant.

$L_1$ distance from the hypersphere chord length distribution can identify that the generated distribution is not uniform and produce a lower-boundary of the error magnitude (i.e. the minimum number of non-uniform points). As a case study, it is assumed that sets $S_{10}$ of $10,000$ $2$-dimensional points are produced using the discussed technique. The (median after $1,000$ runs) $L_1(g_{10})$ distance from $f_2(d)$ is $0.0253$, a value that implies (based on proposition \ref{maximumunfrmsubs}) at least $1,125$ of the $10,000$ points not being generated by a uniform distribution. In comparison, the $1\%$-largest $L_1$ divergence from the theoretic chord length distribution for $10,000$ uniformly selected $2$-dimensional points is $0.0016$, i.e. $16$ times less than the estimated value.

However, in practice, the $1\%$-largest $L_1$ divergence is not expected to be available in such an application, because most of the times this would mean that the person doing the uniformity test already has a second, already debugged, technique generating uniform pointsets. If this isn't true, a different approach is required, one that doesn't need access to the $1\%$-largest $L_1$ divergence. In this case, instead of comparing the $L_1$ distance with some uniformity threshold, we repeat the estimation taking into account only half of the input dataset (i.e. sets $S_5$ of $5,000$ $2$-dimensional points). The corresponding (median after $1,000$ runs) $L_1(g_5)$ distance from $f_2(d)$ is $0.0261$, i.e. only $3.16\%$ higher than $g_1(S_{10})$. As explained in Section \ref{sec:point_uniform}, the convergence rate of uniform distributions to the corresponding hypersphere chord length distribution is $k^{-1/2}$, where $k$ is the number of point-pairs, i.e. approximately $M^{-1}$ where $M$ is the pointset size. Therefore, the expected $L_1(g_{5})$/$L_1(g_10)$ rate is approximately $2$, which is far from the reported $1.0316$, thus signifying a non-uniform spatial distribution. Moreover, the estimated lower boundary of non-uniform points in the set is $11.25\%$ (as already mentioned, the actual value is $21.46\%$).

In general, a simple process to validate algorithms supposedly generating uniform pointsets on the hypersphere is to start by sets of $M_{initial}$ points (e.g.  $M_{initial} = 1,000$) and then iteratively double the pointset size while confirming that the $L_1$ distance ratio of adjacent pointsets is approximately equal to $2$. The process is terminated either when the $L_1$ becomes lower than a uniformity threshold (e.g. $0.001$), in which case the algorithm is validated, or when a $L_1$ distance ratio of adjacent pointsets is near to $1$ (which implies that $L_1$ converges to a non-zero value), in which case the presence of a bug is reported. Such a process could also be applied on the hyper-hemisphere, without requiring any modifications except from the fact that the additional feature of estimating a non-uniformity lower boundary is not available.

\subsection{Evaluating data uniformity}
\label{subsec:evaluating}

The main difference of this setup from the previous one is that a debugged algorithm for generating uniform pointsets on the hypersphere is available and the focus is to assess the uniformity of an input dataset. Such an application would be of great interest in cases where the uniformity (or non-uniformity) of the data is correlated with semantic information about the (partially unknown) process that generated them. Because this definition is too generic, a case study is used to underline the analysis framework, as well as its merit. The employed case study is the spatial distribution of craters on the Moon. 

The population and spatial distribution of Moon craters is of great scientific interest because these quantitative features are related with the age \cite{WKHartmann01} as well as the composition of the Moon surface \cite{MLFeuvre08}. Apart from locally-focused ''crater counting'' \cite{TKneissl15}, analysis of the global features of their distribution has been extensively conducted, including examining their uniformity. As a matter of fact, there is a consensus among planetary scientists that the crater distribution of the Moon (as well as Earth, Mars, etc.) is not uniform. This non-uniformity has been associated with several physical properties such as the latitudinal dependence on the impact velocity and the impact angle \cite{MLFeuvre08}, the angular distance from the apex \cite{TMorota05} and the orbital and size distribution of asteroids and comets in the inner Solar System \cite{MLFeuvre11}.

In \cite{MLFeuvre11} an elaborate quantitative analysis of Moon crater spatial distribution was conducted, including  uniformity assessment. With the use of spherical harmonics \cite{TMMacRobert48} the authors have estimated that the crater rate locally varies from $80\%$ to $125\%$ of the global average. This implies that the Moon crater distribution is a mixture of $80\%$ uniform and $20\%$ non-uniform points defined on a $3$-dimensional sphere. On the other hand, the use of spherical harmonics has revealed no significant difference in uniformity for different crater sizes, based on the maximum/minimum cratering ratio \cite{MLFeuvre11}. Moreover, the authors have reported a symmetry between the North and the South hemisphere. 

In this work, we re-examine \cite{MLFeuvre11} conclusions using the hypersphere chord length distribution. The input data originates from Salamuniccar et al. \cite{GSalamuniccar14}, which introduced the LU78287GT dataset, the most complete lunar crater catalogue that includes the complete list of $22,402$ Moon craters with diameter larger than $8$ kilometres. The coordinates of these $22,402$ craters (which is named set $C$ in the rest of this section) were used to assess crater uniformity. Even though craters of smaller dimensions are available (e.g. LU78287GT consists of $78,287$ craters in total \cite{GSalamuniccar14}) these were ignored because the list is not complete and it can not be undoubtedly assumed that the missing craters do not tamper the uniformity measure.

After computing the crater distance distribution $g_C$ it was compared to $f_3(d)$. The estimated $L_1(g)$ distance was $0.079$, while the $1\%$-largest and the median $L_1$ distance for a uniform $3$-dimensional set of the same size was $5.8~10^{-4}$ and $4.8~10^{-4}$ respectively. Therefore, the chord length distribution uniformity measure confirms that the lunar craters are not uniformly distributed. As a matter of fact, proposition \ref{maximumunfrmsubs} implies that the maximum percentage of uniformly distributed craters is $80.13\%$, an estimate almost identical to the (estimated with spherical harmonics) uniformity reported in \cite{MLFeuvre11}.

Perhaps more interesting is the fact that, contrary to the techniques employed in \cite{MLFeuvre11} (i.e. spherical harmonics and maximum/minimum cratering ratio), using sphere chord length distribution, it is possible to detect size-based uniformity differences. More specifically, the $L_1(g_{>20})$ distance of the distribution of craters larger than $20km$ is $0.0506$ (the $1\%$-largest $L_1$ distance for a uniform set of this size was $1.7~10^{-3}$) while $L_1(g_{<20})$ distance of the distance distribution of craters smaller than $20km$ is $0.1014$ (the $1\%$-largest $L_1$ distance for a uniform set of this size was $9~10^{-4}$). While for the time being there is no theoretic explanation of the root cause of this divergence, this is possibly connected to the fact that (as suggested in \cite{MLFeuvre11}) a distinct numerical model is optimal for the distribution of craters of size larger/smaller than $20km$.

\begin{figure}[htb]
\centering
\includegraphics[width=0.5\textwidth]{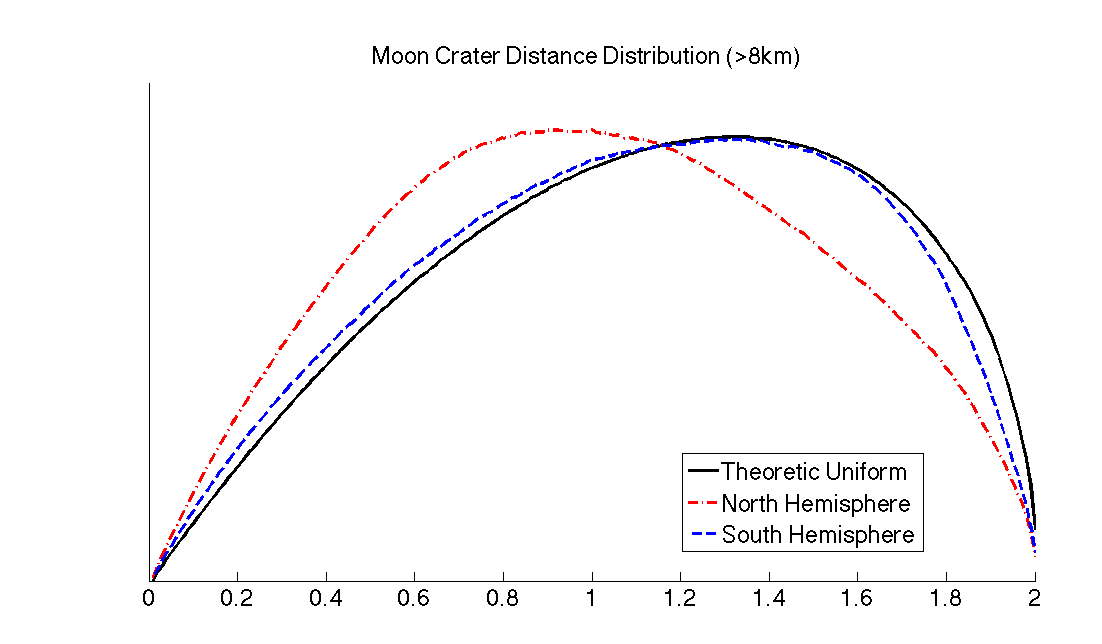}
\caption {The distance distribution of the craters of the lunar North and South hemispheres, compared to the theoretic uniform distribution. While the South hemisphere craters are rather uniformly distributed, in the North hemisphere there seems to be a large divergence from uniformity.}
\label{fig:moon_fig}
\end{figure}

Finally, the crater distance distributions of the North and the South Hemisphere were estimated and compared to the uniform hemisphere chord length distribution. As can be seen in Fig. \ref{fig:moon_fig}, the uniformity-related difference is apparent. The estimated $L_1$ distance is $0.2217$ for the North Hemisphere and $0.047$ for the South Hemisphere (the $1\%$-largest $L_1$ distance for a uniform set of this size was approximately $3~10^{-3}$ in both cases). While for the time being it is not easy to quantify the semantics of this divergence (especially since proposition \ref{maximumunfrmsubs} does not stand for hyper-hemispheres) it is rather straightforward to conclude that there is a difference in uniformity between the two hemispheres, which should be further examined in the future.

In summary, this analysis provides evidence that the introduced hypersphere chord length distribution can contribute in uniformity-related data analysis. Its main advantage over established methods such as spherical harmonics or even simple, grid-based, binning is that it is not based on symmetry or on numeric equivalence (i.e. bins of equal size expected to have an equal number of points) but on a more generic uniformity feature, i.e. the inner structure of a hyperspherical uniform pointset. As a result, the hypersphere chord length distribution can identify subtle non-uniformity instances that are missed by spherical harmonics or/and simple statistics. Since its implementation presents no difficulties and its computational complexity being quadratic is rarely prohibitive, hypersphere chord length distribution constitutes a valuable addition to the tools used to assess data uniformity.

\subsection{Identifying and discarding non-informational data}
\label{subsec:cleaning}

In continuation to the previous sub-section, the maximum size of a uniform subset, which is estimated through the hypersphere chord length distribution, can be used to discriminate the uniform/non-uniform parts of spatial data. This may be of great importance in applications where most of the data are not interesting (e.g. in anomaly detection \cite{JKittler14}).  On the state-of-the-art approach, the key hypothesis is that there is a descriptor space in which the ''interesting data'' (whatever this means) would constitute a compact and clearly defined (i.e. not overlapping with the ''not-interesting'' datsaset) area that can be modelled through some supervised learning technique. Notwithstanding the significant achievements in this kind of applications, there is an inherent theoretical problem with its key hypothesis: if the negative training set represents a \emph{semantically null} set, it is expected to be featureless, therefore not possible to be accurately modelled by some set of descriptors.

An alternative hypothesis would be that a dataset can be projected to a descriptor space as a uniform pointset if-f it is semantically null. Note that in this case the algorithmic focus would shift from making the positive set as descriptive as possible to making the negative set as featureless as possible. If this is correct, then the informational data can be detected by identifying and discarding the uniform background. Such an approach would face two challenges; firstly, to develop this type of descriptors; secondly, to successfully discriminate uniform from non-uniform distribution subsets. 

In this work, it is demonstrated that the second challenge can be met, even by the simple Algorithm \ref{alg:uniformsubsetestim} that is described in Section \ref{sec:point_uniform}. Algorithm \ref{alg:uniformsubsetestim} employs a Monte Carlo nearest neighbour technique in which a number of uniform sets are constructed and projected (using nearest neighbour) onto the mixture of uniform/non-uniform pointsets. The main idea is that the points belonging to the uniform subset will generally have larger support region that the points belonging to the non-uniform one, hence, a ''randomly'' (i.e. uniformly) selected point on the hypersphere would be more probable to have as a nearest neighbour a point in the uniform subset. On the other hand, the $L_1$ distance from the hypersphere chord length distribution implies a maximum size of the uniform set. The Monte Carlo nearest neighbour approach is used as a stochastic estimation of the support region size that is computationally efficient even in high dimensions.

The detection accuracy depends on the span, the shape and the (relative) size of the non-uniform subset, as well as the precision of the threshold implied by the $L_1$ distance. An exhaustive analysis of this approach is not possible due to space limitations. Instead, a rather simple setup is employed focusing on the size of the non-uniform subset and selecting fixed values for the rest of the parameters.

More specifically, a set of $2,000$ $N$-dimensional ($4 \leq N \leq 12$) uniform points represent the non-informational points. Subsequently, an area equal to the $5\%$ of the N-sphere is augmented with more (informational) points so as to finally reach $X\%$ of the total points. Three different values of $X$ are examined, $X=10\%, 15\%$ and $20\%$. The distance distribution of the augmented set $S$ is estimated and its $L_1$ distance from the N-sphere chord length estimation determines (using proposition \ref{maximumunfrmsubs}) the number $M_u$ of uniform points on the augmented set. Finally, $1,000,000$ points are uniformly generated on the N-sphere and projected on their nearest neighbour in $S$. The $M_u$ points with the most points projected on them are discarded and the rest constitute the estimated non-uniform subset. The process was iterated $100$ times for each (N,X) pair and the evaluation is conducted using the (average over the $100$ simulations) Precision-Recall measures. The results are presented in Table \ref{tab:prec_recall_identif_uniform}.

\begin{table}[htb] \scriptsize
\begin{center}
\begin{tabular}{|c||c|c||c|c||c|c|}
\hline N & P. ($10\%$) & R. ($10\%$) & P. ($15\%$) & R. ($15\%$) & P. ($20\%$) & R. ($20\%)$ \\ \hline
4 & 0.4359 & 0.5122 & 0.7210 & 0.5898 & 0.8795 & 0.7101 \\ \hline
5 & 0.5001 & 0.5759 & 0.7946 & 0.6647 & 0.9336 & 0.7372  \\ \hline
6 & 0.5285 & 0.573 & 0.8189 & 0.614 & 0.9489 & 0.7395  \\ \hline
7 & 0.5087 & 0.5374 & 0.8401 & 0.6651 & 0.9545 & 0.7297  \\ \hline
8 & 0.5033 & 0.5136 & 0.8523 & 0.6366 & 0.9583 & 0.7102  \\ \hline
9 & 0.5122 & 0.5304 & 0.8329 & 0.6077 & 0.946 & 0.6744  \\ \hline
10 & 0.5308 & 0.5492 & 0.8501 & 0.6336 & 0.9536 & 0.6786  \\ \hline
11 & 0.5329 & 0.5490 & 0.8430 & 0.5749 & 0.9639 & 0.6663  \\ \hline
12 & 0.5151 & 0.5084 & 0.8351 & 0.5972 & 0.9506 & 0.6358  \\ \hline
\end{tabular}
\end{center}
\caption{Precision and Recall rates of the informational (i.e. non-uniform) point estimation using the setup described in subsection \ref{subsec:cleaning}.} \label{tab:prec_recall_identif_uniform}
\end{table}
The Recall rate is determined from the accuracy of the used threshold, i.e. from how strong the non-uniformity lower boundary of proposition \ref{maximumunfrmsubs} is. Even though the Recall does not exceed $75\%$ in any case, and it seems to fluctuate (and perhaps decrease) when the dimension increases, still a substantial number of non-uniform points is retrieved (in all but one cases, more than $50\%$ of them). Moreover, the Recall is increasing with the size of the non-uniform subset. This can be explained by the fact that proposition \ref{maximumunfrmsubs} makes use of the inequality $L_1(g_c) \leq 2$. The equality $L_1(g_c)=2$ stands if-f $S_c$ (i.e. the non-uniform subset) is a set of identical points (for which the distance distribution is $1$ for zero-distance and $0$ for any non-zero-distance), i.e. if the support region of the non-uniform points is minimum. This implies that the inequality is stronger if the non-uniform points are dense, i.e. if the support region of the non-uniform points is smaller.

On the other hand, the Precision rate increases both with the dimension and with the size of the non-uniform subset (the size of the support region is in this case the main reason for the increase), reaching as high as $96.39\%$. The high Precision rate indicates that the Monte Carlo approach can successfully model the support region size with low computational cost independently from $N$ (at least in the setup examined in this work). This signifies that such an approach is realistically applicable in several different uniform/non-uniform data identification scenarios. Applications that require a higher Recall rate may benefit from the fact that in the employed experimental setup, the center of the estimated non-uniform subset was lying within the ''non-uniform region'' in $87\%$, $92\%$ and $98\%$ of the times for $X$ equal to $10\%, 15\%$ and $20\%$, respectively. Such a property can be used as a basis for the development of more elaborate non-uniform subset estimation algorithms (e.g. through region expansion).  

It is highly possible that  more powerful techniques can be developed using the chord length distribution as a basis. However, neither this task nor the comparison with other state-of-the-art approaches (e.g. mean shift \cite{DComaniciu02}) is within the scope of this work. On the contrary, the analysis objective was to establish that the chord length distribution is potentially useful in a uniform/non-uniform subset detection pipeline, and the evidence provided in this subsection confirms this hypothesis.

\subsection{Uniform sub-set detection embedded in higher dimensional data}
\label{subsec:subset_detection}

The three first experimental sub-sections expanded on the quantitative properties introduced in Section \ref{sec:point_uniform}. The last experimental analysis examines the detection of uniform pointsets in higher dimensions, as discussed in Section \ref{sec:higher_dimensions}. The objective of this section is double; firstly to confirm and quantify the qualitative conclusions driven in Section \ref{sec:higher_dimensions}; secondly, similarly to the other evaluation setups, to give evidence that the hypersphere chord length distribution could be useful in such an application.

In the employed setup, uniform subset detection employs $4$ parameters: (a) the dimension $N_u$ of the uniform subset $S_u$ ($2 \leq N_u \leq 11$) (b) the dimension $N$ of the superset $S$ ($N_u < N \leq 12$) (c) the type $T$ of the superset $S$ ($T=\{Sp,He\}$, where $Sp$ stands for uniform distribution on a hypersphere and $He$ stands for uniform distribution on a hemi-hypersphere) and (d) the ratio $M_u/M$ ($M_u/M = 0.05i, i=\{1,2,...19\}$. Moreover, $M$ was selected to be equal to $5,000$ and $100$ simulations were conducted with each parameter combination.

Once again, the experimental process was designed to be as simple as possible. More specifically, initially the distance matrix of $S$ was estimated, before the distance distribution of each point $p, p \in S$ was estimated and compared with the $N_u$-sphere chord length distribution to estimate the $L_1$ distance. The points with the $M_u$ lowest $L_1$ values were returned and compared with the points of $S_u$ to estimate the detection rate (as a result, in this setup, the ''detection rate'' is equal both to the Recall and to the Precision rate).The average detection rate over the $100$ simulations is reported. Note that this process is not an evaluation scheme that can be used in practice because it assumes that $M_u$ and $N_u$ are a priori known, which is not generally correct. This process focus on evaluating whether the distance distributions $g_{iS}$ from each point $p_i$ have a potential to detect uniform sets in higher dimensions. Optimising the use of $g_{iS}$ distributions in a relevant algorithm is not a work to be done before this potential has become apparent.

Because the parameter space is $4$-dimensional and includes $2,090$ parameter combinations, the results are averaged and compared according to $4$ distinct criteria, each one examining a separate performance factor: (a) the detection rate when the superset $S$ is defined on a hypersphere versus the detection rate when the superset $S$ is defined on a hemi-hypersphere (b) the detection rate as a function of the dimension difference $N-N_u$ (c) the detection rate as a function of $N$ and (d) the detection rate as a function of $M_u/M$.

The average detection rate of the $1,045$ parameter combinations for which $T=Sp$ (i.e. the superset is defined on a hypersphere) is $0.8209$ while the corresponding statistic for the $1,045$ parameter combinations for which $T=He$ (i.e. the superset is defined on a hyper-hemisphere) is $0.8744$. Both rates are significantly better than the baseline of $0.5$, which corresponds to the detection rate if the $M_u$ returned points were randomly selected. Therefore, a first conclusion is that the similarity of the pointwise distance distributions $g_{iS}$ to the hypersphere chord length distribution can be used as a local feature that models uniformity. 

Moreover, there is an apparent difference between $T=Sp$ and $T=He$ runs, which is further confirmed by the fact that there is no run for which the $T=Sp$ detection rate is higher than the corresponding $T=He$ detection rate, while for $21.05\%$ of the runs the $T=He$ detection rate is more than $10\%$ better than the corresponding $T=Sp$ detection rate. This difference is explained by the fact that, as discussed in Section \ref{sec:higher_dimensions}, the detection rate is large when the distance distribution of the superset $S$ is substantially different from the embedded set $S_u$. In the examined setup, if $N_u \approx N$ the $N_u$-sphere distribution is quite similar with the $N$-sphere distribution but not with the $N$-hemisphere distribution. Therefore the detection rate of $T=He$ is substantially higher than the $T=Sp$ one. On the contrary, when $N \gg N_u$ the two detection rates are expected to be quite similar. 

\begin{figure}[htb]
\centering
\includegraphics[width=0.5\textwidth]{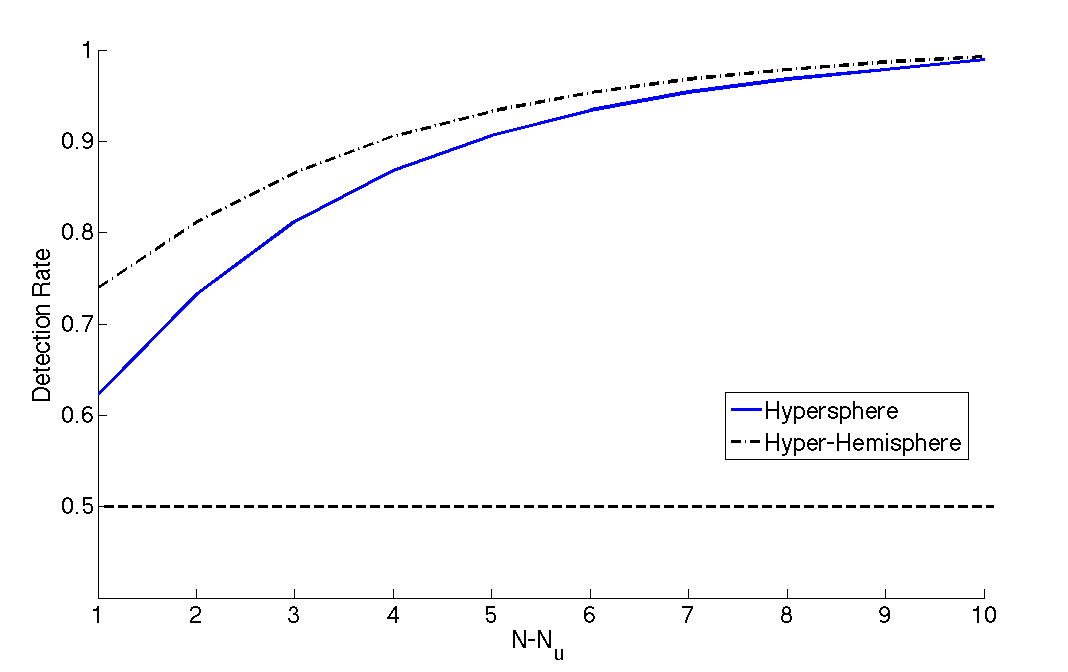}
\caption {The detection rate as a function of $N-N_u$ for $S$ being defined on a hypersphere and on a hyper-hemipshere. The dashed line represents the baseline (i.e. for random selection).}
\label{fig:distance_dimensions}
\end{figure}

This analysis may be further confirmed by the experimental data. For example, while the average detection rate for $(T=He, N_u=2, N=9)$ is only $0.95\%$ higher than the detection rate for $(T=Sp, N_u=2, N=9)$, the average detection rate for $(T=He, N_u=8, N=9)$ is $13.94\%$ higher than the detection rate for $(T=Sp, N_u=8, N=9)$. For a little bit more thorough evaluation the detection rate as a function of $N-N_u$ is plotted in Fig. \ref{fig:distance_dimensions}, showing the hypersphere and the hyper-hemisphere curves to converge for large $N-N_u$. Nevertheless, in general it is easier to  detect uniform subsets when $N_u \ll N$, regardless of the superset distance distribution. However, even for $N-N_u=1$ the detection rate is much better than the (random) baseline, thus verifying the uniformity detection potential of the hypersphere chord length distribution.

\begin{figure}[htb]
\centering
\includegraphics[width=0.5\textwidth]{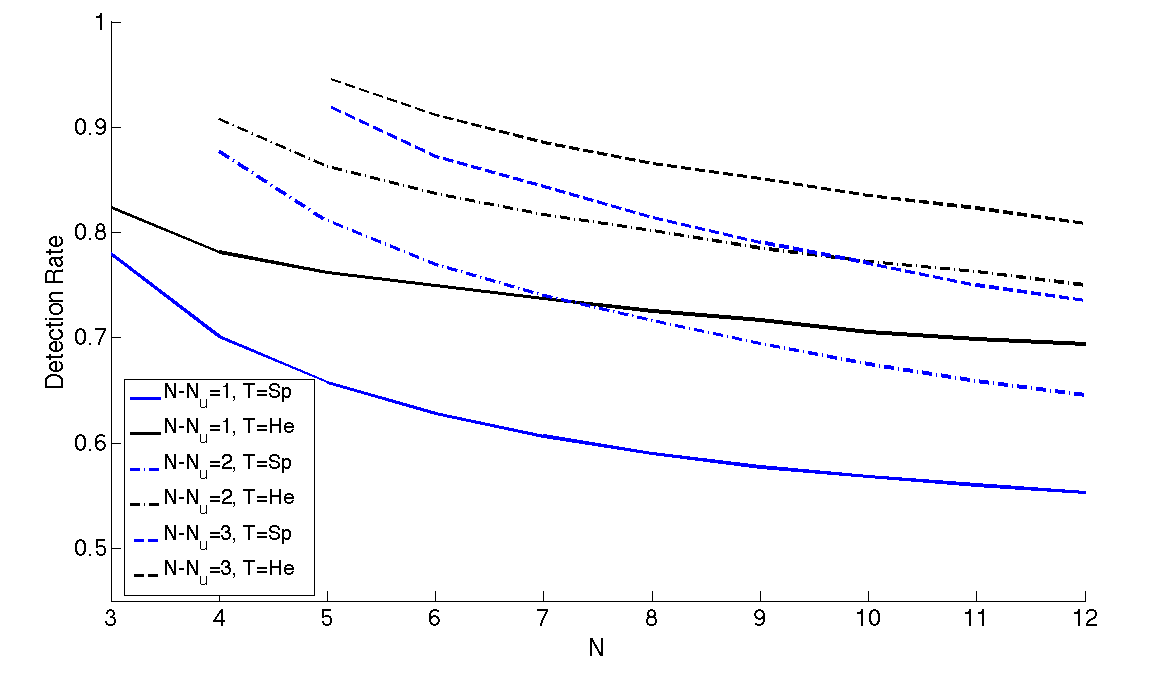}
\caption {The detection rate as a function of $N$ for different $N-N_u$. The blue lines correspond to hypersphere detection rates while the black to hyper-hemispheres. Lines of different colour but same style show detection rate curves for the same $N-N_u$ but for different superset type (hemisphere/hyper-hemisphere).}
\label{fig:distance_N}
\end{figure}

On the other hand, since both the hypersphere and the hyper-hemisphere chord length distribution converge to $\delta (\sqrt2R)$, the uniformity detection potential for a fixed $N-N_u$ is expected to decrease with $N$. In order to examine how fast the performance decline, the detection rate as a function of $N$ for $N-N_u=c, c=\{1,2,3\}$ is plotted (Fig. \ref{fig:distance_N}). Fig. \ref{fig:distance_N} show that apart from the $N-N_u=1, T=Sp$ curve, all other curves are being rather robust in the plotted $N$ range. Moreover, if a $55\%$ detection rate is selected as a low boundary under which the hypersphere chord length distribution is so weak that is practically performing similarly to the baseline, by extrapolating the curves of Fig. \ref{fig:distance_N} it is estimated that for the hypersphere the $N$ that for which the detection rate is below this boundary is $N_{off} = 13+6(N-N_u-1)$ (i.e. if $N-N_u=1$, $N_{off}=13$, if $N-N_u=2$, $N_{off}=19$, if $N-N_u=3$, $N_{off}=25$, etc.) while for the hyper-hemisphere it is $N_{off} = 42+9(N-N_u-1)$. Even though this extrapolation is by default of limited accuracy, it still validates that the examined performance decrease is not prohibitive for a wide range of $N$ and $N_u$ values.

\begin{figure}[htb]
\centering
\includegraphics[width=0.5\textwidth]{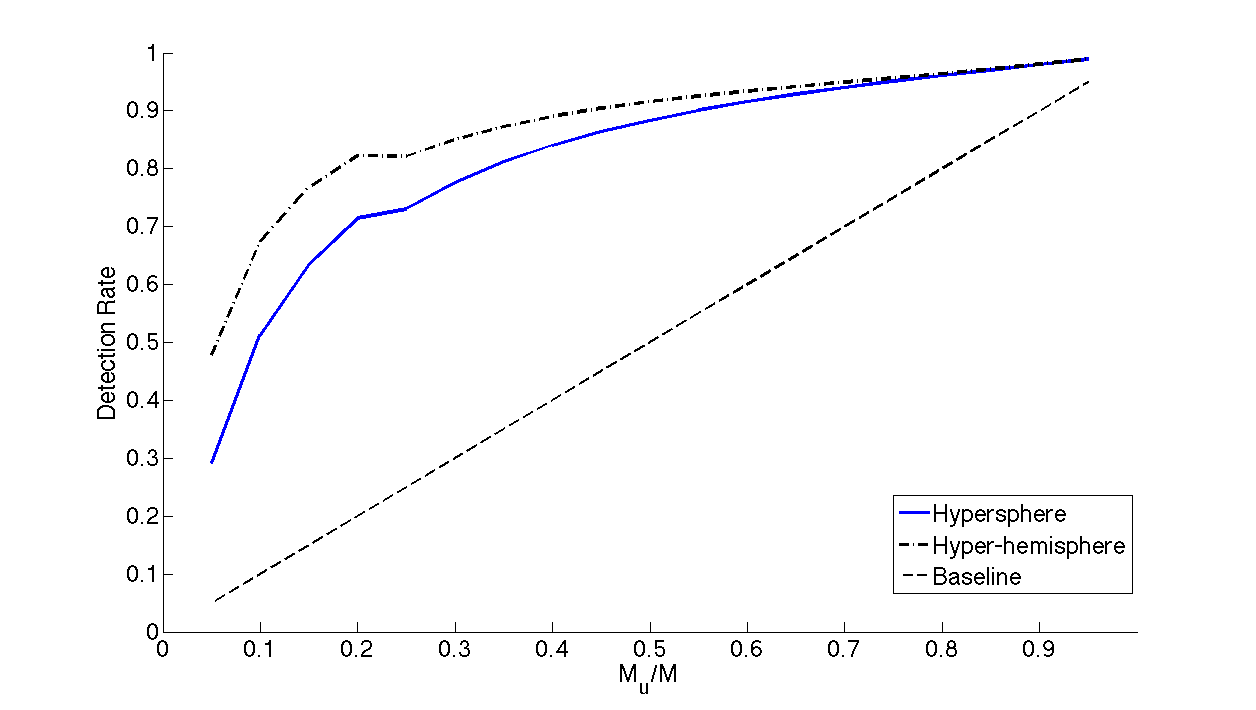}
\caption {The detection rate as a function of $M_u/M$. The black dashed line shows the baseline (i.e. for random selection).}
\label{fig:distance_MuM}
\end{figure}

Finally, the relative size of the uniform subset $M_u/M$ is also related to the detection performance (Section \ref{sec:higher_dimensions}). Typically, the larger the uniform sub-set the higher the performance. However, as shown in Fig. \ref{fig:distance_MuM} the performance increase is far from linear; instead, the detection performance improves rapidly with $M_u/M$ for small $M_u/M$ values and is saturated near to $1$ for large $M_u/M$ values. For example, the $T=Sp$ curve exceeds $0.9$ for $M_u/M=0.55$ while the $T=He$ for $M_u/M=0.45$. Perhaps more importantly, the detection rate is not near the baseline for small uniform sub-sets. For example, the $M_u/M=0.05$ value for the $T=He$ curve is $0.479$, i.e. almost $10$ times better than the baseline. Taking into account that this performance was achieved with the simplest of algorithms it can be deduced that the accurate detection of uniform subsets embedded in higher-dimension data using the hypersphere chord length distribution is possible even for small-sized subsets.

\section{Conclusions and Future Work}
\label{sec:conclusions}

In this work the hypersphere chord length distribution (and the hyper-hemisphere chord length distribution) was analytically introduced and examined, especially in relation to the uniformity of high-dimensional data defined on a hypersphere. Both the theoretic presentation and the experimental evaluation show that the introduced tools can find several applications assessing the uniformity of data. In the future three main directions will be explored.

Firstly, despite its novelty and its potential, the new uniformity measure suffers from being a single-value ''uniformity descriptor''. Notwithstanding its compactness, it is understood that it could greatly benefit from an extension to a vector defined on an orthogonal basis. Theoretically, there is no reason for not being possible to describe a pointset as an infinite sum of uniform distributions on continuously smaller regions (if this was achieved then the similarity to the hypersphere chord length distribution would be just the first term of the infinite sum). As a matter of fact, the main motivation for estimating the hyper-hemisphere chord length distribution was to examine whether this (or a translated/scaled version of it) is orthogonal to the hypersphere chord length distribution. Proposition \ref{propositionhemi1} implies a complex and lengthy expression that is difficult to incorporate in a basis function scheme even for small dimensions. Moreover, the estimation process signify that the chord length distribution of half of the hyper-hemisphere (or even smaller segments of the hypersphere) would be even more complex and impractical to use. Therefore, the extension to an orthogonal basis of gradually more confined uniform distributions does not seem to be achievable by continuously splitting the N-sphere in $2^i, 1 \leq i \leq N$ equal-sized parts. Different possibilities are currently explored that include not only progressively splitting the hypersphere but also updating the distance distribution.

Secondly, it would be useful to have a similar measure for histogram-type of vector data. Histograms is a type of vector data that are extensively used; along with the $L_2$-normalised data (which are defined on a hypersphere) are the most common data types. Histogram variables are non-negative and have a $L_1$ norm equal to $1$, therefore they are not defined on a hypersphere but on a $(N-1)$-dimensional simplex. The distance distribution for uniformly selected points on a high-dimensional simplex, which is currently explored, would allow uniformity measures for histograms to be developed.

Thirdly, the development of algorithms that build upon the measures defined in this work is an ongoing process that is done on an as-needed basis.

\ifCLASSOPTIONcaptionsoff
  \newpage
\fi



\bibliographystyle{IEEEtran}
\bibliography{references}
%
%
%

%

%

\end{document}